\documentclass[aps,twocolumn,showpacs,superscriptaddress,amsmath,amssymb]{revtex4-1}

\usepackage{graphicx}
\usepackage{color}
\usepackage{ulem}

\newcommand{\eqreff}[1]{Eq.~(\ref{#1})}
\newcommand{\Eqreff}[1]{Eq.~(\ref{#1})}

\newcommand{\subj}{\rm {act}}
\newcommand{\obj}{\rm {pass}}
\begin{document}

\title{
Complementary aspects of non-equilibrium thermodynamics
}

\author{Lee Jinwoo}
\email{e-mail: jinwoolee@kw.ac.kr}
\affiliation{Department of Mathematics, Kwangwoon 
University, 20 Kwangwoon-ro, Nowon-gu, Seoul 01897, Korea}

\author{Hajime Tanaka}
\email{e-mail: tanaka@iis.u-tokyo.ac.jp}
\affiliation{Department of Fundamental Engineering, Institute of Industrial Science, University of Tokyo, 4-6-1 Komaba, Meguro-ku, Tokyo 153-8505, Japan}

\date{\today}

\begin{abstract}
{\bf 
Bio-molecules are active agents in that they consume energy to perform tasks. The standard theoretical description, however, considers only a system-external work agent. Fluctuation theorems, for example, do not allow work-exchange between fluctuating molecules. This tradition leaves `action through work', an essential characteristic of an active agent, out of proper thermodynamic consideration. Here, we investigate thermodynamics that considers internal-work. We find a complementary set of relations that capture the production of free energy in molecular interactions while obeying the second law of thermodynamics. This thermodynamic description is in stark contrast to the traditional one. A choice of an axiom whether one treats a portion of Hamiltonian as `internal-work' or `internal-energy' decides which of the two complementary descriptions manifests among the dual. We illustrate, by examining an allosteric transition and a single-molecule fluorescence-resonance-energy-transfer measurement of proteins, that the complementary set is useful in identifying work content by experimental and numerical observation.
}
\end{abstract}

\maketitle

The bio-molecular process is an essential class of chemical reactions, involving big molecules which fold \cite{anfinsen1975}, change their configurations \cite{shape}, and associate and dissociate with each other or with chemicals \cite{molecular,ham2012pnas} to carry out their biological functions. The energy landscape theory \cite{funnel1991,funnel1992,funnel1997} has successfully explained those molecular phenomena using non-equilibrium statistical mechanics \cite{revSears2008, jarReview, revSeifert}, taking into account fluctuations at the single-molecule level \cite{liph2001,expColin,ritort2012}. The theory, for example, explains molecules' personalities \cite{personality2007} or molecule-to-molecule variations \cite{hyeon2012} by effective-ergodicity-breaking transitions over a long observation time \cite{roldan2014}, caused mainly due to the ruggedness of the free energy landscape \cite{multiple2010}.  

T. Hill pioneered thermodynamic formalism of molecular phenomena for a steady-state assuming detailed balance and time-scale separation between slow variables (e.g., molecular machines) and fast variables (e.g., ATP molecules) \cite{hill1962,hill2012}. It enables us to discuss what type of thermodynamic changes have occurred to biological systems, and study the budget of work and entropy production with thermodynamic rigor \cite{parmeggiani1999,fisher1999, fisher2001, bustamante2001, schmiedl2008, liepelt2007,hwang2016,hwang2018}.

Beyond steady-states, the study of non-equilibrium thermodynamics is still in progress. Hatano and Sasa have extended the second-law of thermodynamics for transitions between steady states \cite{sasa}. Seifert has discussed the second-law along a single trajectory for non-equilibrium processes in terms of stochastic entropy that is time-dependent and converges to equilibrium entropy as a system goes to equilibrium \cite{seifert05, revSeifert}.
Bochkov and Kuzovlev (BK) have introduced work to the formalism of fluctuation theorems \cite{bochkov1977,bochkov1979}. Jarzynski and Crooks have linked non-equilibrium work to equilibrium free energy \cite{jar,crooks99}. The authors have related work to point-wise non-equilibrium free energy \cite{local}.

Depending on whether work affects the Hamiltonian of a system or not, thermodynamic descriptions of a non-equilibrium process become fundamentally different. In Jarzynski's equality, external perturbation $\lambda_t$ modifies the Hamiltonian and thus equilibrium free energy, deforming the energy landscape itself of a system, while it does not in BK's approach, merely driving a system out of equilibrium on a fixed energy landscape \cite{jar2007work,campisi2011}. 
In both cases, $\lambda_t$ that mediates work-exchange (see Fig. 1A) should vary in a pre-determined and repeatable manner \cite{liph2001,liph2002, expSasa, expColin,ritort2012}. Thus if a work source fluctuates (e.g., cases as depicted in Fig. 1B), work fluctuation theorems do not apply. 

In this paper, we consider a situation where system-internal perturbation (e.g., the interaction between molecules) drives a non-equilibrium process. We take two different approaches: One is that the interaction between interacting molecules affects the Hamiltonian and thus the equilibrium free energy of a system (see Fig. 1C), and the other is that it does not modify the Hamiltonian but merely drives them out of equilibrium (see Fig. 1D). We refer the former and the latter to views I and II, respectively. We will prove a work fluctuation theorem in terms of a new quantity of $\Psi$ that encodes work for both cases and derive complementary sets of thermodynamic relations for a non-equilibrium process.

\begin{figure} \label{fig:fig1}
\includegraphics[width=8cm]{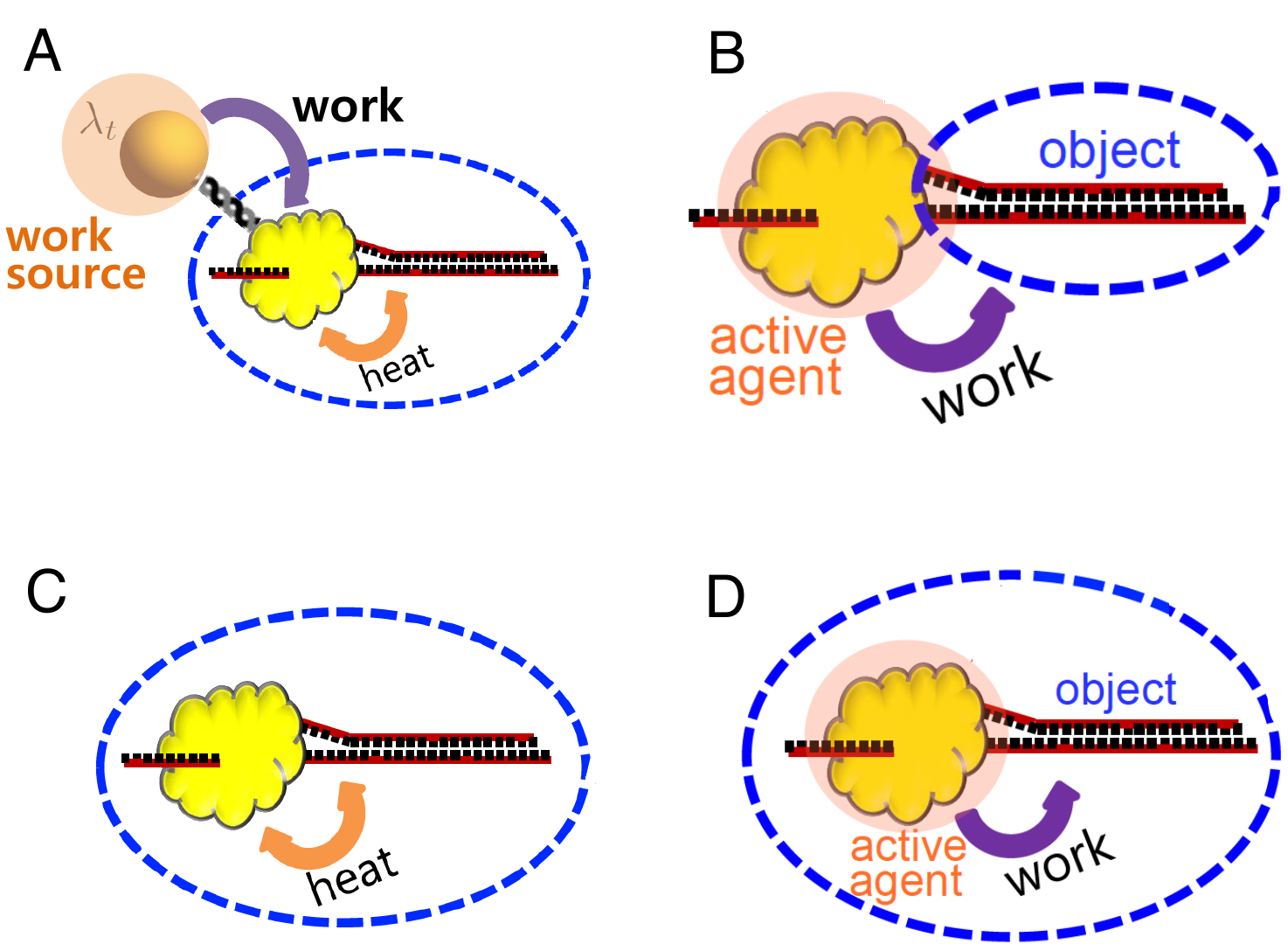}
\caption{{\bf Work in fluctuation theorems.} 
A molecular system $z=(x,y)$ composed of helicase ($x$; yellow object) and DNA ($y$; red-black zipper) is schematically represented.
In each case, the shaded area in light pink is the location of a work source, and the blue dashed ellipse indicates the boundary of a system, which is defined by the independent variable of non-equilibrium free energy $\psi$ as we describe below.
{\bf A,} External control $\lambda_t$ that varies in a pre-determined manner applies work to the molecular system $z$. The system includes molecules $x$ and $y$ and is described by $\psi(z; \lambda_t)$.
{\bf B,} If a work source (helicase) fluctuates, work fluctuation theorems do not apply. Depicted boundary of the system indicates that we are considering $\psi(y,\cdot)$, including DNA only, upon which helicase $(x)$ works.
{\bf C,} The interaction between the two molecules is being considered as a part of internal-energy, which we call view I.
{\bf D,} The interaction between the two molecules is being considered as internal-work, which we call view II.
In {\bf (C, D)}, we consider $\psi(z,\cdot)$ so that the system includes both molecules. 
It is crucial to note the fundamental difference between {\bf B} and {\bf D} in the system's boundary: In the former, $x$ works on $y$, but in the latter, $z$ works on itself.
}
\end{figure}

\section{Results}
\subsection{A state function $\Psi$ that encodes work}
We decompose the phase space of a system into disjoint mesoscopic states $\{\chi_j | j=1,\cdots,J\}$. We also partition the time axis $\{\tau_k|k=0,\cdots,K\}$ with $\tau_0=0$, depending on the time resolution ($\delta \tau_k$) of an experiment. 
An unusual aspect of our approach is to treat an intermediate non-equilibrium state $\chi_j$ at time $\tau_k$ as a thermodynamic ensemble, more specifically, an ensemble of paths to the state $\chi_j$ from an initial probability distribution \cite{local}, which enables us to treat $\chi_j$ with thermodynamic rigor (see Fig. 2).

Let $\lambda_t$ for $0\le t \le \tau$ be an arbitrary process which we repeat with an initial probability distribution $p(\chi_j,\tau_0)$. 
We define $\Psi$ for each mesoscopic state $\chi_j$ at coarse-grained time $\tau_k$ as follows:
\begin{eqnarray}\label{eq:psi}
\Psi(\chi_j,\tau_k):=-\beta^{-1}\ln\left<e^{-\beta\psi(z,t)}\right>_{z\in\chi_j,t\in\tau_k},
\end{eqnarray}
where $\psi(z,t)$ is local free energy of microstate $z$ at time $t$, $\beta$ is the inverse temperature ($\beta := 1/(k_{\rm B}T)$, where $k_{\rm B}$ is the Boltzmann constant and $T$ is the temperature of the heat bath), and the brackets indicate the average over all $z\in\chi_j$ and $t\in\tau_k$ with respect to the conditional probability $p(z,t | \chi_j,\tau_k)$. We remark that $\Psi$ is an original quantity that is different from the average non-equilibrium free energy since we have $\Psi(\chi_j,\tau_k)\le \left<\psi(z,t)\right>_{z\in\chi_j,t\in\tau_k}$, where the equality holds if and only if local equilibrium holds. 



Local free energy $\psi$ and thus $\Psi$ depend on internal energy, which differs in the two approaches (views I and II) that we take. Thus we will discuss them in detail below. Temporarily, we indicate the dependency on internal energy by subscript ${\rm v}$.
Let $G_{\rm v}(\chi_j; \tau_k)$ be the (time-averaged) conformational free energy of $\chi_j$, i.e. $G_{\rm v}(\chi_j; \tau_k):=-\beta^{-1}\ln\left< \int_{z\in\chi_j}e^{-\beta E_{\rm v}(z; \lambda_t)} dz \right>_{\tau_k}$, where $E_{\rm v}$ is system's internal energy, and $\left<\cdot\right>_{\tau_k}$ indicates the average over time $t\in\tau_k$. We note that $G_{\rm v}(\chi_j; \tau_k)$ is well defined regardless of whether the condition of local equilibrium within $\chi_j$ holds or not. 
Now \eqreff{eq:psi} implies
\begin{equation}\label{eq:prob2}
p(\chi_j,\tau_k)=\frac{e^{-\beta G_{\rm v}
(\chi_j; \tau_k)}}{e^{-\beta \Psi _{\rm v}(\chi_j,\tau_k)}},
\end{equation}
which holds in full non-equilibrium situations (see Appendix A.1 for the proof).
If we further assume the microscopic reversibility \cite{kur, maes1999,crooks99,jar2000}, we have
\begin{equation}\label{eq:psi2}
\Psi_{\rm v}(\chi_j,\tau_k) =  -\beta^{-1}
\ln \left<e^{-\beta W_{\rm tot}} \right>_{\chi_j},
\end{equation}
where $W_{\rm tot}:=W_{\rm v} + \psi_{\rm v}(z,0)$, $W_{\rm v}$ is work, and the brackets indicate the average over all paths to state $\chi_j$ at time $\tau_k$ (see Appendix A.2 for the proof). \eqreff{eq:psi2} tells that $\Psi_{\rm v}(\chi_j,\tau_k)$ encodes a property of the ensemble of trajectories that reach $\chi_j$ at time $\tau_k$ from an initial probability distribution, more specifically, an amount of work done for reaching state $\chi_j$ at time $\tau_k$. %
\eqreff{eq:psi2} implies a refined version of the second law of thermodynamics within each ensemble $\chi_j$:
\begin{equation}\label{eq:2nd-law}
\left<W_{\rm v}\right>_{\chi_j}\ge \Delta\Psi_{\rm v}(\chi_j,\tau_k),
\end{equation}
where $\Delta\Psi_{\rm v}(\chi_j,\tau_k):= \Psi_{\rm v}(\chi_j,\tau_k)-\left<\psi_{\rm v}(z,0)\right>$ . Here the last term, $\left<\psi_{\rm v}(z,0)\right>$, is the average non-equilibrium free energy at time $0$ (see Appendix A.2 for the proof).

\begin{figure}[h]
\label{fig:fig2}
\includegraphics[width=7cm]{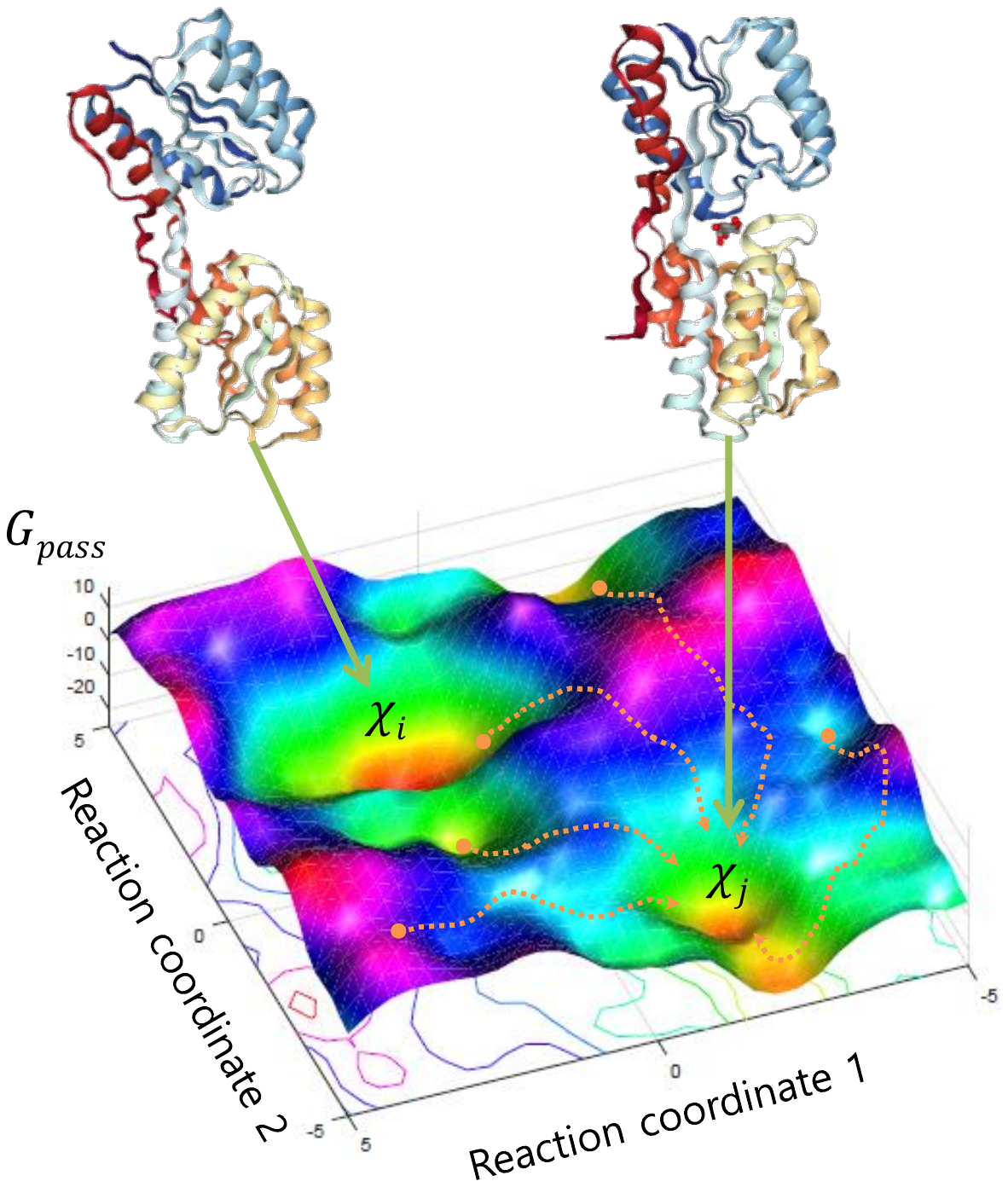}
\caption{
{\bf Non-equilibrium state $\chi_j$ as an ensemble of trajectories.}
The colored surface is a schematic representation of the free energy landscape $G_{\obj}$ of a molecular system in terms of hypothetical reaction coordinates 1 and 2. The molecular states $\chi_i$ and $\chi_j$ are crystal structures of open and closed conformations of a ribose-binding protein, respectively \cite{pdb2003}. Some members of barrier-overcoming trajectories reaching the closed conformation $\chi_j$ are presented in the orange dotted curved arrows. Each trajectory consumes work that is internally supplied through the interaction with a ribose. An ensemble average of these work along all paths to $\chi_j$ at time $\tau$ from an initial distribution $p_0(\chi_j$) determines $\Psi_{\rm II}(\chi_j,\tau)$ (see \eqreff{eq:psi2}). 
}
\end{figure}

\subsection{A choice of an axiom; internal-work vs. internal-energy}
Let us consider that a finite classical stochastic system weakly coupled to the heat bath of inverse temperature $\beta$ is composed of two subsystems. 
We may consider an active agent to act upon another molecule (see Figs. 1C and 1D). Let $z=(x,y)$, where $x (y)$ denotes a phase space point of each molecule. We set $\lambda_t$ as follows: 
We assume that before time $\tau_0$, two subsystems are in inert equilibrium with Hamiltonian 
\begin{equation}\label{eq:pass}
E_{\obj}(z):=E_x(x)+E_y(y),
\end{equation}
where $E_x$ and $E_y$ are Hamiltonians of each subsystem. At time $\tau_0$, they start to interact through activating interaction energy $E_{\rm int}(x,y)$:
\begin{equation}\label{eq:act}
E_{\subj}(z):= E_{\rm int}(x,y).
\end{equation}
To motivate two different approaches that we take, we start by obtaining the energy balance equation for this process following Jarzynski's approach \cite{jar}.

We may describe the change of internal energy of the system during the process as follows:
\begin{equation}\label{eq:E1}
E(z, t):=E_{\obj}(z)+H(t)E_{\subj}(z),
\end{equation} 
where $H(t)$ is the Heaviside step function that is $H(t):=1$ for $t>0$ and $H(t):=0$ for $t\le 0$, and the time derivative of $H(t)$ gives the Dirac delta function $\delta(t)$. 
Then Jarzynski's work $W_{\rm jar}$ done on the system along trajectory $\{z_t\}_{0\le t\le\tau}$ is 
\begin{equation}\label{eq:jar_w}
W_{\rm jar}:=\int_0^\tau \frac{\partial E}{\partial t}\,dt = \int_0^\tau \delta(t)E_{\subj}(z_t)\,dt = E_{\subj}(z_0),
\end{equation}
which is the interaction energy evaluated at the initial point $z_0$ \cite{jar}.
Jarzynski equation gives the difference between the initial equilibrium free energy $A_0$ and the final one $A_1$ in terms of $W_{\rm jar}$, telling that the equilibrium free energy instantly changes due to the activated interaction energy.

Now, \eqreff{eq:E1} and \eqreff{eq:jar_w} provide the following energy balance equation:
\begin{equation}\label{eq:jar}
E_{\subj}(z_0) = \Delta E_{\obj} + E_{\subj}(z_\tau) + Q_b,
\end{equation}
where $\Delta E_{\obj}:=E_{\obj}(z_\tau)-E_{\obj}(z_0)$, and $Q_b$ is dissipated heat.
We would like to rewrite \eqreff{eq:jar} into the following canonical form of the first law of thermodynamics:
\begin{equation}\label{eq:view0}
W_{\rm v} = \Delta E_{\rm v} + Q_b,
\end{equation} 
where $W_{\rm v}$ is work done on the system, $\Delta E_{\rm v}$ is the change of an internal energy, and subscript ${\rm v}$ indicates the dependency on our different approaches.
A comparison of \eqreff{eq:jar} to \eqreff{eq:view0} gives us two options; moving $E_{\subj}(z_0)$ in \eqreff{eq:jar} to the right-hand side or moving $E_{\subj}(z_\tau)$ to the left-hand side.

If we move $E_{\subj}(z_0)$ in \eqreff{eq:jar} to the right-hand side, 
the first law of thermodynamics reads
\begin{equation}\label{eq:view1}
0 = \Delta E + Q_b,
\end{equation}
where $\Delta E := \Delta E_{\obj} + \Delta E_{\subj}$, and $\Delta E_{\subj}:=E_{\subj}(z_\tau)-E_{\subj}(z_0)$. From \eqreff{eq:pass} and \eqreff{eq:act}, $E(z) = E_x(x)+E_{\rm int}(x,y)+E_y(y)$ is the total energy of the composite system. The local free energy then becomes
\begin{equation}\label{eq:small1}
\psi_{\rm I}(z,t):=E(z)-\beta^{-1}\sigma(z,t),
\end{equation}
where $\sigma(z,t):=-\ln p(z,t)$ is the stochastic entropy \cite{crooks99,seifert05}. 
We define $\Psi_{\rm I}(\chi_j,\tau_k)$ by \eqreff{eq:psi} accordingly (see Appendix A.3 and A.4). Here subscript I indicates this first approach (v=I), say view I (see Fig. 1C).

If we move $E_{\subj}(z_\tau)$ in \eqreff{eq:jar} to the left-hand side and set
\begin{equation}\label{eq:2}
W_{\subj} := -\Delta E_{\subj},
\end{equation}
the first-law of thermodynamics reads
\begin{equation}\label{eq:view2}
W_{\subj}  = \Delta E_{\obj}+Q_b.
\end{equation}
Contrary to view I, the corresponding local free energy is subtle:
In \eqreff{eq:2}, the conformational change of molecules secretes some energy $-\Delta E_{\subj}$, which we treat as internal-work $W_{\subj}$. If we took that portion of energy not only as internal-work but also as a part of internal-energy, it would violate the first law of thermodynamics, \eqreff{eq:view2}, by counting double the same portion of energy. Thus, we are forced to have
\begin{equation}\label{eq:small2}
\psi_{\rm II}(z,t):=E_{\obj}(z)-\beta^{-1}\sigma(z,t).
\end{equation}
We define $\Psi_{\rm II}(\chi_j,\tau_k)$ by \eqreff{eq:psi} accordingly (see Appendix A.3 and A.4). Here subscript II indicates this second approach (v=II), say view II (see Fig. 1D). It is important not to confuse this case, where the system includes both $x$ and $y$, with that depicted in Fig. 1B, where the system includes only $y$ (cf.~\cite{balian,jar2013prx}, where a work source as a subsystem does not fluctuate).

We remark three things: Firstly, views II and I respectively treat the activated energy $E_{\subj}(z)$ as internal-work and internal-energy, which behave in a complementary manner: if one increases during a process, the other decreases, forming a precursor of a drastic divergence of thermodynamics of the process as we describe below. 
Secondly, except tradition, no physical constraint forces us to select either one of the two approaches, making a choice axiomatic.
Thirdly, \eqreff{eq:2} can deal with generalized (e.g., entropic) forces by including relevant degrees of freedom involved in the effects (e.g., energy from chemical potentials by incorporating some solution molecules into the system \cite{seifert2011stochastic}).

\subsection{A simple process in the two views}

\begin{figure*}[ht]
\label{fig:fig3}
\includegraphics[width=16cm]{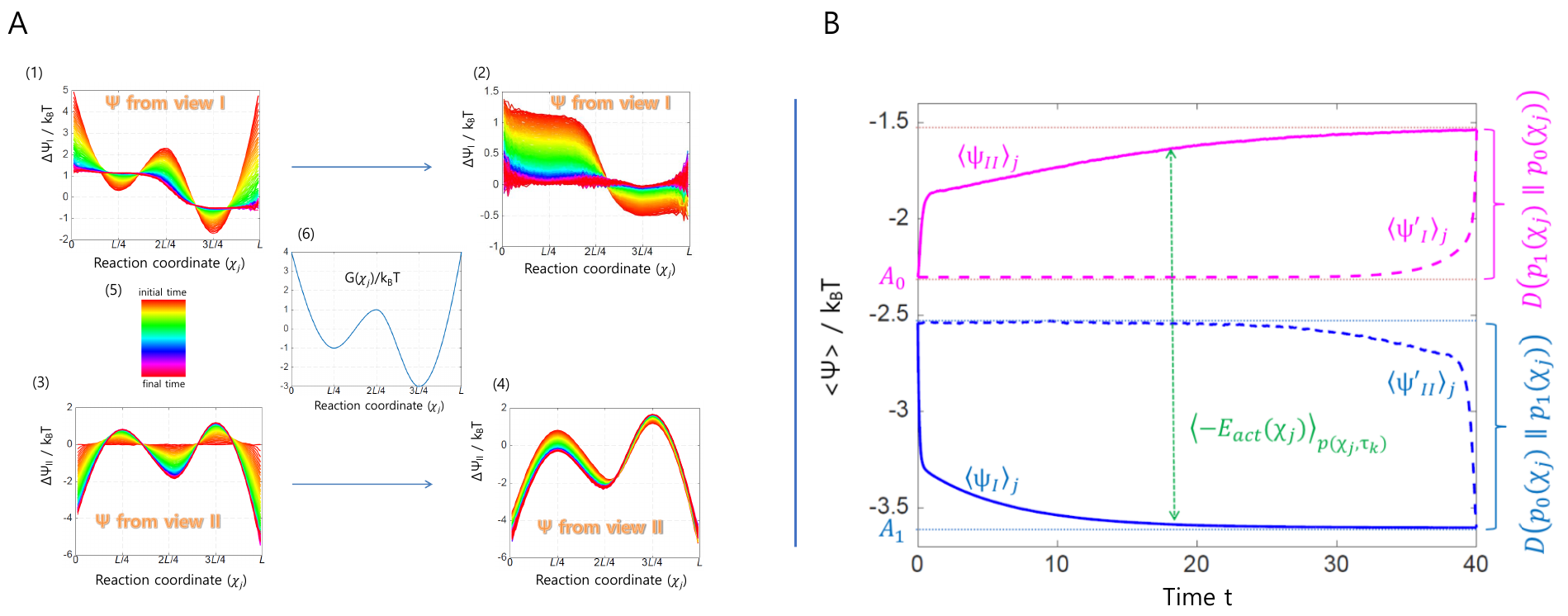}
\caption{
{\bf The time evolution of $\Psi(\chi_j,\tau_k)$ and its average $\left<\Psi\right>_j$ from views I and II.}
{\bf A(1)-(2),} $\Delta \Psi_{\rm I}(\chi_j,\cdot):=\Psi_{\rm I}(\chi_j,\cdot) - A_1$, where $A_1$ is the final equilibrium free energy, indicates the extent to which each state $\chi_j$ is away from the final equilibrium state. {\bf (1)} shows $\Delta\Psi_{\rm I}(\chi_j,\tau_k)$ from $\tau_0$ to $\tau_{100}$ and {\bf (2)} that from $\tau_{101}$ to $\tau_{5000}$. 
{\bf A(3)-(4),}  $\Delta \Psi_{\rm II}(\chi_j,\tau_k):=\Psi_{\rm II}(\chi_j,\tau_k)-A_0$, where $A_0$ is the initial equilibrium free energy, indicates the amount of work that the ensemble of molecules has generated to form $\chi_j$ at time $\tau_k$ from an initial equilibrium state. {\bf (3)} shows $\Delta\Psi_{\rm II}(\chi_j,\tau_k)$ from $\tau_0$ to $\tau_{100}$ and {\bf (4)} that from $\tau_{101}$ to $\tau_{5000}$.
{\bf A(5),} The color code for (1)-(4) is shown. 
{\bf A(6),} The conformational free energy $G(\chi_j)$ for a hypothetical one dimensional reaction coordinate $\chi_j$ is shown. 
{\bf B,} The solid pink curve shows $\left<\Psi_{\rm II}\right>_j$, the average of $\Psi_{\rm II}$ over all $\chi_j$, for $\tau_0< t < \tau_{4000}$, and the solid blue curve is for $\left<\Psi_{\rm I}\right>_j$. The dotted curves are those quantities for the reverse process. $E_{\subj}(\chi_j):=-\beta^{-1}\ln\left<e^{-\beta E_{\subj}(z)}\right>_{p_0(z|\chi_j)}$, where the brackets indicate the average over all $z\in\chi_j$ with respect to $p_0(z|\chi_j)$, is the instant Jarzynski's work, and $D$ indicates the Kullback-Leiber divergence. $\left<-E_{\subj}(\chi_j)\right>_{p(\chi_j,\tau_k)}$ indicates the average of $-E_{\subj}(\chi_j)$ over all $\chi_j$ with respect to $p(\chi_j,\tau_k)$.
}
\end{figure*}

Let us consider that a molecular system is in equilibrium in a hypothetical one-dimensional reaction coordinate $\chi_j$ $(0\le\chi_j\le L)$ (see Appendix B.1 on the details on the simulation). Here we set the initial probability distribution $p_0(\chi_j)$ of the system's state to be in $\chi_j$ at time $\tau_0=0$ to be uniform:
\begin{equation}\label{eq:p0}
p_0(\chi_j)\propto e^{-\beta G_{\obj}(\chi_j)}=|\chi_j|,
\end{equation}
where $G_{\obj}(\chi_j):=-\beta^{-1}\ln\int_{z\in\chi_j}e^{-\beta E_{\obj}(z)} dz$ with $E_{\obj}(z):=0$ under the assumption that the volume $|\chi_j|$ is constant for all $j$.
At time $\tau_0$, one activates energy $E_{\subj}(z)$ such that the conformational free energy $G(\chi_j) \left(:=-\beta^{-1}\ln\int_{z\in\chi_j} e^{-\beta E(z)} dz \right)$ is as depicted in Fig. 3A(6). 
After the activation at time $\tau_0$, the system would evolve towards a new steady-state:
\begin{equation}\label{eq:p1}
p_1(\chi_j)\propto e^{-\beta G(\chi_j)}. 
\end{equation}
We assume a linear relationship between forces and fluxes so that the Fokker-Plank equation governs the dynamics of the probability $p(\chi_j,\tau_k)$ of finding the system at state $\chi_j$ at time $\tau_k$ \cite{mesoscopic,qian2001relative}. 
Fig. 3 shows the time evolution of this {\it bare} process in terms of $\Psi_{\rm I}$ (Figs. 3A(1)-(2)) and $\Psi_{\rm II}$ (Figs. 3A(3)-(4)).

In view I, the initial equilibrium distribution $p_0(\chi_j)$ suddenly becomes a non-equilibrium one due to the activated energy $E_{\subj}(z)$. In detail, the work fluctuation theorem, \eqreff{eq:psi2}, applied at time $\tau_0$ using Jarzynski's work $W_{\rm jar}$ in \eqreff{eq:jar_w} together with energy balance \eqreff{eq:jar} gives $\Psi_{\rm I}(\chi_j,\tau_0)-A_0$, which is the amount of abrupt change from the initial equilibrium free energy $A_0$ due to the activated energy $E_{\subj}(z)$ (see Appendix A.3 for the proof). 
For time $\tau_k>\tau_0$, the deviation of the probability $p(\chi_j,\tau_k)$ from the final probability $p_1(\chi_j)$ would cause the dynamics of $p$. \eqreff{eq:prob2} implies
\begin{equation}\label{eq:G1}
\Psi_{\rm I}(\chi_j,\tau_k) = A_1 + \beta^{-1}\ln\left[ \frac{p(\chi_j,\tau_k)}{p_1(\chi_j)}\right],
\end{equation}
where $A_1 \left(:=-\beta^{-1}\ln\sum_{\chi_j}\int_{z\in\chi_j} e^{-\beta E(z)} dz \right)$ is the final equilibrium free energy. Thus, $\Delta \Psi_{\rm I} := \Psi_{\rm I}(\chi_j,\tau_k)-A_1$ in Figs. 3A(1)-(2) indicates the extent to which each state $\chi_j$ is away from the final state, and the imbalance of $\Psi_{\rm I}(\chi_j,\tau_k)$ in $\chi_j$ leads the destination of the ensemble of paths to the steepest descent direction of $\Psi_{\rm I}$ until the ensemble resolves the imbalance: $\Psi_{\rm I}(\chi_j,\tau_\infty)=A_1$ for all $\chi_j$ \cite{local}. 
Taking average of \eqreff{eq:G1} with respect to $p(\chi_j,\tau_k)$ over all $\chi_j$ gives
$\left<\Psi_ {\rm I}(\chi_j,\tau_k)\right>_{j} = 
A_1 + D
(p(\chi_j,\tau_k)\parallel p_1(\chi_j))$,
where $D(p \parallel p_1)$ is the Kullback-Leibler divergence which is positive and converges to $0$ as $p$ approaches to $p_1$ (see the solid blue curve in Fig. 3B).
Thus, we have 
\begin{equation}\nonumber
\left<\Psi_{\rm I}\right>_j \mbox{tends to be minimized}.
\end{equation}
In short, view I defines thermodynamics of the process as the relaxation of a non-equilibrium towards the equilibrium state.

In view II, \eqreff{eq:prob2} implies
\begin{equation}\label{eq:G2}
\Psi_{\rm II}(\chi_j,\tau_k) = A_0 + \beta^{-1}\ln \left[ \frac{p(\chi_j,\tau_k)}{p_0(\chi_j)}\right],
\end{equation}
where $A_0 \left(:=-\beta^{-1}\ln\sum_{\chi_j}\int_{z\in\chi_j} e^{-\beta E_{\obj}(z)} dz \right)$ is the initial equilibrium free energy. The effect of the activation of $E_{\subj}(z)$ is not immediate; it does not change neither the internal energy of the system, \eqreff{eq:small2}, nor $\Psi_{\rm II}$ at time $\tau_0$ in \eqreff{eq:G2}. 
As the molecule changes its conformation such that it generates power as internal-work as in \eqreff{eq:2}, it increases $\Psi_{\rm II}$ as \eqreff{eq:psi2} indicates. Thus $\Delta\Psi_{\rm II}(\chi_j,\tau_k):= \Psi_{\rm II}(\chi_j,\tau_k)-A_0$ in Figs. 3A(3)-(4) shows the amount of internal work that the ensemble of molecules has generated to form $\chi_j$ at time $\tau_k$. 
Taking average of \eqreff{eq:G2} over $\chi_j$ gives
$\left<\Psi_ {\rm II}(\chi_j,\tau_k)\right>_{j} = 
A_0 + D
(p(\chi_j,\tau_k)\parallel p_0(\chi_j))$,
where $D(p \parallel p_0)$ becomes larger as $p$ continues to deviate from $p_0$ as the system-internal work accumulates (see the solid pink curve in Fig. 3B).
Thus, we have (see also Appendix A.5)
\begin{equation}\nonumber
\left<\Psi_{\rm II}\right>_j \mbox{tends to be maximized}.
\end{equation}
We note that this statement does not violate the second law of thermodynamics which reads as follows:
\begin{equation}\label{eq:avg-2nd-law}
\left<W_{\subj}\right>\ge\left<\Delta\Psi_{\rm II}\right>_j.
\end{equation}
\eqreff{eq:avg-2nd-law} holds from \eqreff{eq:2nd-law} by taking average over all $\chi_j$.

In summary, view II defines the thermodynamics of the process as the accumulation of internal-work that drives the system away from equilibrium. We note that \eqreff{eq:G1} and \eqreff{eq:G2} hold in full non-equilibrium situations with no time-scale separation assumption between fast ($z\in\chi_j$) and slow ($\chi_j$) variables (see Appendix A.4 for additional relations in the two views).

\begin{figure*}[t]
\label{fig:fig4}
\includegraphics[width=16cm]{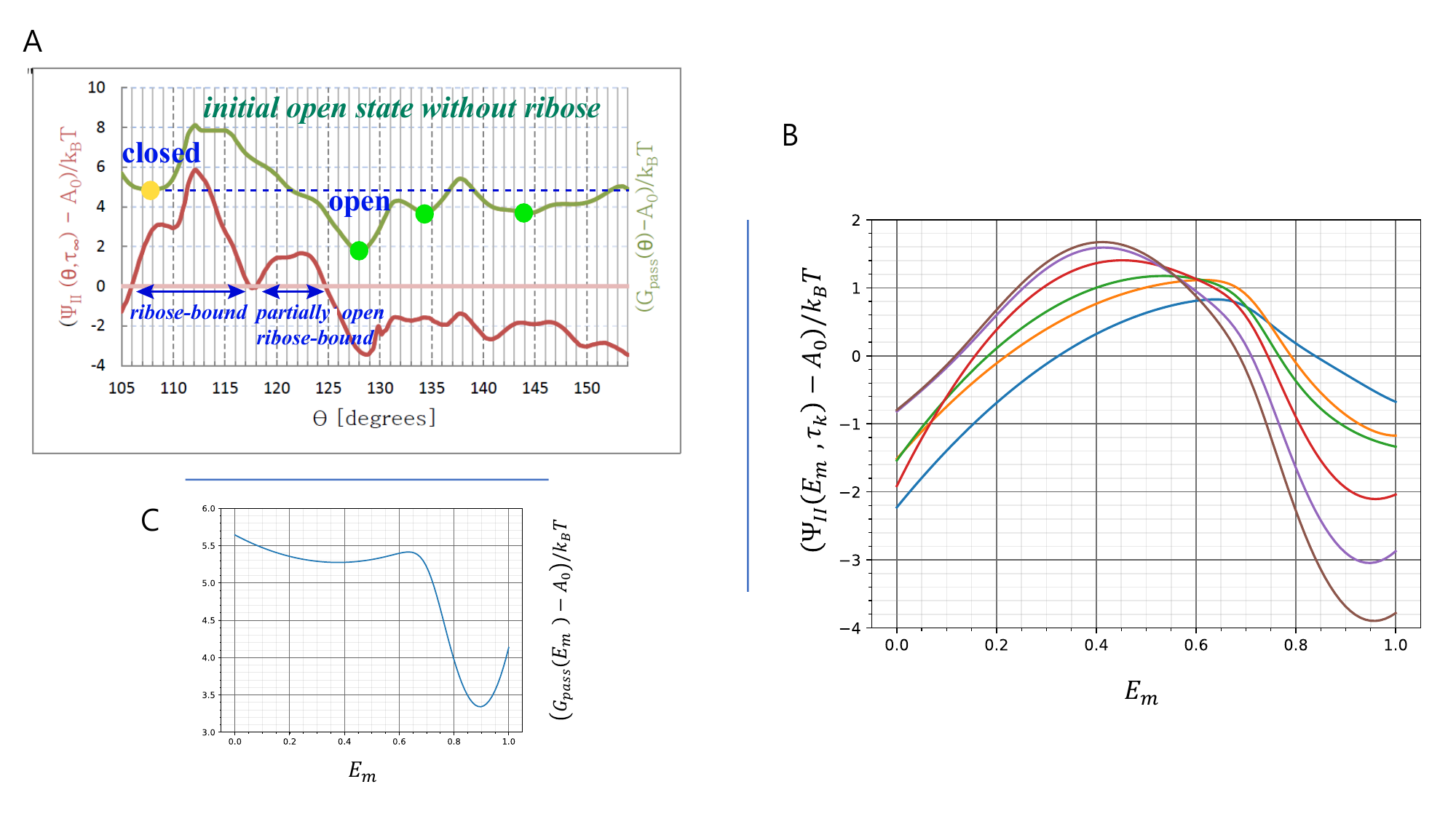}
\caption{
{\bf A work-content revealed from an equilibrium computer simulation of the ribose-binding protein (RBP) and a non-equilibrium kinetic experiment of a microfabricated laminar-flow mixer coupled to the measurement of F\"orster resonance energy transfer (FRET) for the cold-shock protein (Csp).}
{\bf A,} The light green curve is the conformational free energy $G_{\obj}(\theta)$ at the initial ribose-free state and the red curve is $\Psi_{\rm II}(\theta,\tau_{\infty})$ that encodes a work-content for forming the state $\theta$ at time $\tau_\infty$.
{\bf B,} A work-content of each state characterized by the measured FRET efficiency $E_m$ after an abrupt change at time $\tau_0$ of solution conditions to those that favor unfolded states, $E_m\approx 0.4$, from those that favor the folded state, $E_m\approx 0.95$. Blue, orange, green, red, purple, and brown curves represent $\Psi_{\rm II}(E_m, \tau_k)$ at $\tau_k=0.1, 0.2, 0.5, 1.0, 2.0$, and $4.0$ seconds, respectively.
{\bf C,} The conformation free energy $G_{\obj}(E_m)$ of Csp in the initial (before mixing) equilibrium solution conditions  that favor the folded state is shown.      
}
\end{figure*}

\subsection{A work content of each state}
The work content revealed from view II provides more valuable information in realistic cases than in previous artificial ones, as we illustrate below. 

\subsubsection{Ribose-binding protein}
We consider the allosteric transition of ribose-binding protein (RBP), which is an essential class of bio-molecular reactions (see also Appendix B.2).
We set 
\begin{equation*}
\begin{aligned}
E_{\obj}(z) &:= E_{\rm RBP}(x) + E_{\rm rib}(y)\\
E_{\subj}(z)&: = E_{\rm int}(x,y),
\end{aligned}
\end{equation*}
where $E_{\rm RBP}(x)$ and $E_{\rm rib}(y)$ are the internal energy of RBP and ribose, respectively, $E_{\rm int}(z)$ gives interaction between them, and $z=(x,y)$ is a phase-space point of the system.
A reaction coordinate that we consider is angle $\theta$ between the centers of mass of the two domains of RBP and the center of mass of the hinge (see the crystal structures of open and closed conformations in Fig. 2). 

Before time $\tau_0$, the system is in inert equilibrium with Hamiltonian $E_{\obj}(z)$. 
The light green curve in Fig. 4A from computer simulations in \cite{JMB2005} represents the energy landscape 
\begin{equation}
G_{\obj}(\theta):=-\beta^{-1}\ln\left[\int_{z\in\theta} e^{-\beta E_{\obj}(z)} dz\right],
\end{equation} 
where the open state $(\theta\ge 125^\circ)$ is more stable than the closed state $(\theta<125^\circ)$. It indicates that ribose-binding is an active process that should overcome the free energy barrier $G_{\obj}(\theta)$ over $124^\circ \ge \theta \ge 112^\circ$ separating the two states.

At time $\tau_0$, one activates $E_{\subj}(z)$, which we treat as internal-work. Then, the non-equilibrium probability distribution in \eqreff{eq:prob2} reads:
\begin{equation}\label{eq:22}
p(\theta, \tau_k) = \frac{\exp\left[-\beta G_{\obj}(\theta)\right]}{\exp\left[-\beta\Psi_{\rm II}(\theta,\tau_k)\right]},
\end{equation}
which decomposes a molecular state into two contributions; one that forms the free energy barrier and the other that overcomes the barrier to perform a task.
The red curve in Fig. 4A calculated using \eqreff{eq:G2} upon data from \cite{JMB2005} represents a work-content $\Psi_{\rm II}(\chi_j,\tau_\infty)$. Here we define $\Delta \Phi_{\rm II}(\chi_j,\tau_\infty)= \Psi_{\rm II}(\chi_j,\tau_\infty)-A_0$. Then, we can, for example, obtain the following information from Fig.~4A. The ensemble of molecules applies work of $\Delta\Psi_{\rm II}(115^\circ,\tau_\infty)\approx 3 k_{\rm B}T$ on average in the sense of \eqreff{eq:psi2} to reach the right edge of the plateau of $G_{\obj}(\theta)$ at $\theta=115^\circ$ and should spend additional energy of $3 k_{\rm B}T$ to walk over to the left edge, in total, applying internal-work of $\Delta\Psi_{\rm II}(112^\circ,\tau_\infty)\approx 6 k_{\rm B}T$ in reaching $\theta=112^\circ$ for ribose-binding from the initial inert equilibrium state of free energy $A_0$.

We have two remarks: Firstly, $\Delta\Psi_{\rm II}(112^\circ,\tau_\infty)\approx \Delta G_{\obj}(112^\circ) \left(:= G_{\obj}(112^\circ)-G_{\obj}(128^\circ)\right)$, where $G_{\obj}(128^\circ)$ is the conformational free energy of the mostly populated state in the initial ribose-free condition. It seems to be reasonable since \eqreff{eq:2nd-law} tells that $\Delta\Psi_{\rm II}(\chi_j,\tau_\infty)$ forms the minimum of average internal-work $\left<W_{\subj}\right>_{\chi_j}$ for reaching $\chi_j$ from the initial equilibrium ensemble, in which the mostly populated state is dominant.
Secondly, a sharp drop in $\Psi_{\rm II}(\theta,\tau_\infty)$ for $108^\circ\le \theta\le 112^\circ$ indicates that the cause of binding in this regime is not the interaction energy between RBP and ribose (since $-W_{\subj}=\Delta E_{\rm int} > 0$ in this regime), but a structural preference of RBP (and ribose) for a more closed state since $G_{\obj}(108^\circ)<G_{\obj}(112^\circ)$, where $G_{\obj}(\theta)=-\beta^{-1}\ln\int_{(x,y)\in\theta}e^{-\beta (E_{\rm RBP}(x)+E_{\rm rib}(y))}\,dx\,dy$. 

In this way, view II provides us with invaluable intuitive information on the biological reaction.

\subsubsection{Single-molecule FRET measurement}

The new perspective (view II) also enables us to extract from a kinetic experiment of single molecules a work content during molecular interactions under conditions far from equilibrium. We analyze an experiment reported in \cite{lipman2003single} that uses a microfabricated laminar-flow mixer coupled to the measurement of F\"orster resonance energy transfer (FRET) efficiencies of individual protein molecules under an abrupt change of solution conditions (see also Appendix B.3).

In detail, the cold shock protein (Csp) that is labeled with fluorescent donor and acceptor dyes at the terminal Cys residues is under equilibrium in pH 7 phosphate buffer. They triggered unfolding by mixing the solution with a denaturant, 8 M guanidinium chloride (GdmCl). After mixing is completed by the time about 50 ms, FRET efficiencies are measured at chosen regions, which correspond, via a fixed flow rate of their experimental setup, to times $\tau_k\approx$ 0.1 , 0.2 , 0.5 , 1.0 , 2.0 , and 4.0 seconds after triggering unfolding \cite{lipman2003single}.

GdmCl molecules unfold a protein by engaging in hydrogen bonds with the protein backbone or solvating the charged residues of a protein, by directly altering electrostatic interactions \cite{interactions2007}.
Thus we can extract the amount of electrostatic interactions from the kinetic experiment by setting as follows: 
\begin{eqnarray}
\begin{aligned}
E_{\obj}(z)&:=E_{\rm CspInBuffer}(x) + E_{\rm GdmCl}(y)\\
E_{\subj}(z)&:=E_{\rm int}(x,y),
\end{aligned}
\end{eqnarray}
where $E_{\rm CspInBuffer}(x)$ is the internal energy of the solution $(x)$, i.e. Csp and buffer molecules, $E_{\rm GdmCl}(y)$ is the internal energy of GdmCl molecules $(y)$, and $E_{\rm int}(x,y)$ gives the interaction between the solution and GdmCl. We consider that before time $\tau_0$, the solution with inert GdmCl is in equilibrium with Hamiltonian $E_{\obj}(z)$, and at time $\tau_0$, the interaction between the solution and GdmCl, which we treat as internal-work, is turned-on.

A reaction coordinate that we consider is the measured FRET efficiency $E_m := n_a / (n_a+n_d)$ at chosen regions that correspond to the delays $\tau_k$. Here $n_d$ and $n_a$ are the sums (in 30 minutes) of donor counts and acceptor counts, respectively, for each single molecule event that lasts for 1~ms \cite{interactions2007}. We note that a standard correction procedure that is well-established very recently \cite{precision2018} enables one to convert $E_m$ to the distance between fluorophores using the F\"orster theory \cite{roy2008practical,joo2008advances} so that one can count thermodynamic states \cite{schuler2008protein}.

From the relative event probability of measured FRET efficiencies $p(E_m,\tau_k)$ at each time $\tau_k$ \cite{lipman2003single}, we obtained the temporal change of $\Psi_{\rm II}(E_m,\tau_k)$ that encodes the amount of interactions between the solution and GdmCl molecules in forming state $E_m$, as shown in Fig. 4B. Here we used \eqreff{eq:G2} with $p_0(E_m)$, which is the probability of $E_m$ before mixing the denaturant. Fig 4C shows $\Delta G_{\obj}(E_m) \left(:=G_{\obj}(E_m)-A_0\right)$ calculated by $-\beta^{-1}\ln p_0(E_m)$.

We note that $E_m\approx 0.95$ corresponds to the folded state where most of the fluorescence photons are emitted by the acceptor due to such lower dye-dye separation as 1~nm. In such a situation, the excitation of the donor dye by a focused laser results in rapid energy transfer to the acceptor dye \cite{lipman2003single}. On the other hand, lower FRET efficiencies $E_m\approx 0.4$ indicates greater dye-dye separation that corresponds to unfolded states.

After time $\tau_0$, staying the unfolded state requires energy from the heat bath against $W_{\subj}:=-\Delta E_{\subj}$ that favors unfolding. 
Fig. 4B shows that it amounts from $-\Delta\Psi_{\rm II}(0.95,\tau_k)\approx 0.4 k_{\rm B}T$ to $4.0 k_{\rm B}T$ as time flows, which could occur very rarely at last.
Most of interaction energy has been consumed to make Csp unfolded, $E_m\approx 0.4$, where we observe that as time flows, ${\Delta\Psi_{\rm II}(0.4, \tau_k) \longrightarrow \Delta G_{\obj}(0.4):=G_{\obj}(0.4)-G_{\obj}(0.95)}$, which is $1.8k_{\rm B}T$ . 

This example also shows the usefulness of the thermodynamic descriptions based on view II.

\section{Discussion} 

The following equality in \cite{jar_lag2009} is well known 
\begin{equation}\label{eq:jar_lag1}
\frac{p(z,t)}{p^{\rm eq}(z,\lambda_t)}=\frac{e^{-\beta\Delta A}}{ \left<e^{-\beta W(t)}\right>_{z,t}},
\end{equation}
where $W$ is Jarzynski's work, $\Delta A:=A_t-A_0$, and $A_t$ and $p^{\rm eq}(z,\lambda_t)$ are respectively the equilibrium free energy and the equilibrium probability density corresponding to the value of external parameter $\lambda_t$ at time $t$. \Eqreff{eq:jar_lag1} implies
\begin{equation}\label{eq:jar_lag2}
\left<W\right>_{z,t} - \Delta A \ge \beta^{-1}\ln \frac{p(z,t)}{p^{\rm eq}(z,\lambda_t)},
\end{equation}  
which is also well-known \cite{jar_lag2009}.

Technically, we have decomposed \eqreff{eq:jar_lag1} into \eqreff{eq:psi}, \eqreff{eq:prob2}, and \eqreff{eq:psi2} so that \eqreff{eq:jar_lag2} reduces to \eqreff{eq:2nd-law} of view I. 
Considering, for example, molecular interactions as system-external perturbation by setting $W = W_{\rm jar}$ as in \eqreff{eq:jar_w} (so that $\Delta A \neq 0$), \eqreff{eq:jar_lag2} reads by \eqreff{eq:G1}:
\begin{eqnarray}
\begin{aligned}\label{eq:ext}
\left<W_{\rm jar}\right>_{z,t} &\ge \Psi_{\rm I}(z,t)-A_0,
\end{aligned}
\end{eqnarray}
as we have discussed (see Appendix A.3).
Alternatively, with an appropriate non-equilibrium preparation of system's initial state, we may set $\lambda_t:=E(z)$ as in \eqreff{eq:view1} for $t\ge0$ (including $t=0$ so that $\Delta A=0$ like BK's approach), then, \eqreff{eq:jar_lag2} reads by \eqreff{eq:G1}: 
\begin{equation} 
0=\left<W\right>_{z,t} \ge \Psi_{\rm I}(z,t)-A_1.
\end{equation}

Conceptually, we have introduced a new perspective that considers system-internal perturbation as thermodynamic work. View II is similar to BK's approach in the sense that work does not modify a system's Hamiltonian. However, it is fundamentally different from BK's approach as well as conventional approaches, in both of which work is defined as system-external perturbation.
Only with substituting Jarzynski's work by internal-work as in \eqreff{eq:2}, \eqreff{eq:jar_lag2} enables us to extract an internal-work content from molecular interactions using \eqreff{eq:G2}:
\begin{equation}\label{eq:int}
\left<W_{\subj}\right>_{z,t} \ge \Psi_{\rm II} (z,t)-A_0,
\end{equation}  
since we have $\Delta A=0$.

\section{Summary} 
Conventionally, the thermodynamic description of a system has exclusively considered a system-external work agent, or an external viewpoint. Here we have considered how to make a thermodynamic description from the viewpoint of an internal-work agent. We have introduced a new state function $\Psi(\chi_j)$ for mesoscopic state $\chi_j$ with no local equilibrium assumption and linked it to work done by the internal agent. We have derived a complementary set of relations, showing that thermodynamic states seen from an internal-work agent are fundamentally different from those seen from an external viewpoint (compare, for example, \eqreff{eq:ext} and \eqreff{eq:int}). We have demonstrated that the new thermodynamic description based on the system-internal work agent not only provides a fresh perspective on the non-equilibrium evolution of a system but also are useful for quantifying molecules' efforts, or internal-work in chemical and biological reactions.  


\vspace{1cm}

\noindent
{\bf Acknowledgments.}
L.J. thanks Chan-Woong Lee for discussions. L.J. was supported by the National Research Foundation of Korea Grant funded by the Korean Government (NRF-2010-0006733, NRF-2012R1A1A2042932, NRF-2016R1D1A1B02011106). 
H.T. acknowledges support from Grants-in-Aid for Specially Promoted Research (JP25000002) and Scientific Research (A) (JP18H03675) from the Japan Society for the Promotion of Science (JSPS).

\noindent
{\bf Author Contributions.}
L.J and H.T conceived the work, developed the theory,
and wrote the paper.

\noindent
{\bf Additional information.}
Correspondence and requests for materials should be addressed to L.J (jinwoolee@kw.ac.kr) or H.T. (tanaka@iis.u-tokyo.ac.jp). 

\noindent
{\bf Competing financial interests.}
The authors declare no competing financial interests.

\onecolumngrid
\appendix

\section{Proof and analysis}\label{sec4}
We consider a finite classical stochastic system weakly coupled to a heat bath of inverse temperature $\beta:=1/(k_{\rm B}T)$ where $k_B$ is the Boltzmann constant and $T$ is the temperature of the heat bath. 
We decompose the phase space of a system into disjoint mesoscopic states $\{\chi_j | j=1,\cdots,J\}$. We also partition the time axis $\{\tau_k|k=0,\cdots,K\}$ with $\tau_0=0$, depending on the time resolution ($\delta \tau_k$) of an experiment. 

\subsection{Proof of equation (2)} 
Let us define the local form of non-equilibrium free energy \cite{local}:
\begin{equation}\label{eq:s1}
\psi_{\rm v}(z,t) := E_{\rm v}(z; \lambda_t) + \beta^{-1}\ln p(z,t),
\end{equation}
where subscript v indicates viewpoint dependency, $E_{\rm v}(z; \lambda_t)$ is an internal energy of a system at microstate $z$, $\lambda_t$ is external control, and $p(z,t)$ is the probability density of $z$ at time $t$. The definition of $\Psi$ in equation (1) implies  
\begin{eqnarray}\label{eq:s2}
\begin{aligned}
e^{-\beta\Psi_{\rm v}(\chi_j,\tau_k)} &= \frac{1}{p(\chi_j,\tau_k)\delta \tau_k}\int_{z\in\chi_j,t\in\tau_k} e^{-\beta\psi_{\rm v}(z,t)}p(z,t) dz dt \\
&= \frac{1}{p(\chi_j,\tau_k)\delta \tau_k}\int_{z\in\chi_j,t\in\tau_k} e^{-\beta E_{\rm v}(z; \lambda_t)} dz dt\\
&= \frac{e^{-\beta G_{\rm v}(\chi_j; \tau_k)}}{p(\chi_j,\tau_k)}.
\end{aligned}
\end{eqnarray}
Here $G_{\rm v}(\chi_j; \tau_k):=-\frac{1}{\beta}\ln\left<\int_{z\in\chi_j} e^{-\beta E_{\rm v}(z; \lambda_t)} dz\right>_{\tau_k}$, where brackets indicate the average over time $\tau_k$. We used \eqreff{eq:s1} to obtain the second line from the first.
Rearranging \eqreff{eq:s2} with respect to $p(\chi_j,\tau_k)$ gives:
\begin{equation}\label{eq:s3}
p(\chi_j,\tau_k)=\frac{e^{-\beta G_{\rm v}(\chi_j;\tau_k)}}{e^{-\beta\Psi_{\rm v}(\chi_j,\tau_k)}},
\end{equation}
which proves equation (2) in the main text.

\subsection{Proof of equations (3) and (4)}
We (temporarily) consider an arbitrary process $\lambda_t$ for $0\le t\le \tau$ with well-defined initial probability density $p(z,0)$ for each microstate $z$ of a system. We denote the phase-space point of the system at time $t$ as $z_t$. For each trajectory $\{z_t\}_{0\le t\le \tau}$, we define the time-reversed conjugate as $\{z'_t\}_{0\le t\le \tau}:=\{z^*_{\tau-t}\}$, where $*$ denotes momentum reversal. 
We assume that internal energy is invariant upon momentum reversal and assume the microscopic reversibility \cite{kur, maes1999,crooks99,jar2000}:
\begin{equation}\label{eq:s4}
\frac{p(\{z_t\}|z_0)}{p'(\{z'_t\}|z'_0)}=e^{\beta Q_b},
\end{equation}
where $Q_b$ is energy transferred to the heat bath, and $p(\{z_t\}|z_0)$ is the conditional probability of path $\{z_t\}$ given
initial point $z_0$ and $p'(\{z'_t\}|z'_0)$ is that for the reverse process which is defined by $\lambda'_t:= \lambda_{\tau-t}$ for $0\le t\le \tau$ and $p'(z'_0, 0) : = p(z_\tau,\tau)$. Now we have 
\begin{eqnarray}\label{eq:s5}
\begin{aligned}
\frac{p(\{z_t\})}{p'(\{z'_t\})}&=\frac{p(\{z_t\}|z_0)\cdot p(z_0,0)}{p'(\{z'_t\}|z'_0)\cdot p'(z'_0,0)}\\
&=\exp\{\beta Q_b + \beta\Delta E_{\rm v} -\beta \Delta \psi_{\rm v}\}\\
&=\exp\{\beta W_{\rm v} - \beta\Delta \psi_{\rm v}\},
\end{aligned}
\end{eqnarray}
where $\Delta\psi_{\rm v} := \psi_{\rm v}(z_\tau,\tau) - \psi_{\rm v}(z_0,0)$. Here we used the first law of thermodynamics, equation (10), which reads $W_{\rm v} = \Delta E_{\rm v} + Q_b$, and \eqreff{eq:s1}. Focusing on trajectories $\Gamma_{\chi_j,\tau_k}$ that reach $\chi_j$ at time $\tau_k$, \eqreff{eq:s5} implies 
\begin{eqnarray}\label{eq:s6}
\begin{aligned}
\left<e^{-\beta W_{\rm tot}}\right>_{\chi_j} &:= \left<e^{-\beta W_{\rm v}-\beta\psi_{\rm v}(z_0,0)}\right>_{\chi_j}\\
&= \frac{1}{p(\chi_j,\tau_k)\delta\tau_k}\int_{\tau\in\tau_k}\int_{\{z_t\}\in\Gamma_{\chi_j,\tau}} e^{-\beta W_{\rm v}-\beta\psi_{\rm v}(z_0,0)} p(\{z_t\}) d\{z_t\} d\tau\\
&= \frac{1}{p(\chi_j,\tau_k)\delta\tau_k}\int_{\tau\in\tau_k}\int_{\{z_t\}\in\Gamma_{\chi_j,\tau}} e^{-\beta \psi_{\rm v}(z_\tau,\tau)} p'(\{z'_t\}) d\{z_t\} d\tau\\
&= \frac{1}{p(\chi_j,\tau_k)\delta\tau_k}\int_{\tau\in\tau_k}\int_{z_\tau\in\chi_j} e^{-\beta \psi_{\rm v}(z_\tau,\tau)} p(z_\tau,\tau) dz_\tau d\tau\\
&=e^{-\beta\Psi_{\rm v}(\chi_j,\tau_k)},
\end{aligned}
\end{eqnarray}
which proves equation (3) in the text. Here we used $p(z_\tau,\tau) = p'(z'_0, 0)$, and the assumption that internal energy is invariant upon momentum reversal, and $d\{z_t\}=d\{z'_t\}$.
The application of Jensen's inequality to \eqreff{eq:s6} implies a refined version of the second law of thermodynamics (equation (4)):
\begin{equation}
\left<W_{\rm v}\right>_{\chi_j} \ge \Delta\Psi_{\rm v}(\chi_j,\tau_k),
\end{equation}
where $\Delta\Psi_{\rm v}(\chi_j,\tau_k):=\Psi_{\rm v}(\chi_j,\tau_k)-\left<\psi_{\rm v}(z_0,0)\right>$. Here the last term, $\left<\psi_{\rm v}(z_0,0)\right>$, is the average of $\psi_{\rm v}(z_0,0)$ with respect to $p(z_0,0)$.

\subsection{The relationship between the two views and Jarzynski's work}

From now on, we restrict our attention to the following process: Before time $\tau_0$, two subsystems ($x$ and $y$) are in inert equilibrium with Hamiltonian 
\begin{equation}\label{eq:pass}
E_{\obj}(z):=E_x(x)+E_y(y),
\end{equation}
where $E_x$ and $E_y$ are Hamiltonians of each subsystem, and $z=(x,y)$ is a microstate of the total system. At time $\tau_0$, they start interaction by activating interaction energy $E_{\rm int}(x,y)$:
\begin{equation}\label{eq:act}
E_{\subj}(z):= E_{\rm int}(x,y).
\end{equation}
We define $E(z):=E_{\obj}(z)+E_{\subj}(z)$.

Now we take views I and II in combination, instead of taking a single point of view exclusively, and repeat the proof of the work fluctuation theorem, which clarifies the relationship between the two views and the activated energy interpreted as Jarzynski's work.
By combining the energy balance equation (9) and Jarzynski's work, we obtain:
\begin{equation}
W_{\rm jar} = E_{\subj}(z_0) = \Delta E_{\obj} + E_{\subj}(z_\tau) + Q_b,
\end{equation}
together with $p(z_0,0)=\exp\{-E_{\obj}(z_0)+\psi_{\rm II}(z_0,0)\}^\beta$ in \eqreff{eq:s1} from view II, and $p(z_\tau,\tau)=\exp\{-E(z_\tau)+ \psi_{\rm I}(z_\tau,\tau)\}^\beta$ in \eqreff{eq:s1} from view I.
We note that $\Delta E_{\obj} :=E_{\obj}(z_\tau)-E_{\obj}(z_0)$ and $\psi_{\rm II}(z_0,0)=A_0$, where $A_0:=-\beta^{-1}\ln\sum_{\chi_j}\int_{z\in\chi_j}e^{-\beta E_{\obj}(z)} dz$ is the initial equilibrium free energy. Then, \eqreff{eq:s5} reads:
\begin{eqnarray}
\begin{aligned}
\frac{p(\{z_t\})}{p'(\{z'_t\})}&=\exp\{Q_b + \Delta E_{\obj}+ E_{\subj}(z_\tau) - \psi_{\rm I}(z_\tau,\tau) + A_0\}^\beta\\
&=\exp\{ W_{\rm jar} - \psi_{\rm I}(z_\tau,\tau) + A_0\}^\beta,
\end{aligned}
\end{eqnarray}
By following the same argument as in \eqreff{eq:s6}, we obtain
\begin{equation}\label{eq:s12}
\Psi_{\rm I}(\chi_j,\tau_k) = A_0 - \beta^{-1}\ln\left<e^{-\beta W_{\rm jar}}\right>_{\chi_j}.
\end{equation}
At time $\tau_0=0$, \eqreff{eq:s12} tells that Jarzynski's work gives $\Psi_{\rm I}(\chi_j,0)-A_0$, which is the amount of abrupt change from the initial equilibrium free energy $A_0$ due to the activated energy $E_{\subj}(z_0)$.

We may rewrite \eqreff{eq:s12} at time $\tau_0=0$ as follows:
\begin{equation}\label{eq:s13}
  \Psi_{\rm I}(\chi_j,0)-\Psi_{\rm II}(\chi_j,0)= E_{\subj}(\chi_j),
\end{equation}
where 
\begin{equation}\label{eq:s14}
E_{\subj}(\chi_j):=-\beta^{-1}\ln\left<e^{-\beta E_{\subj}(z_0)}\right>_{p_0(z_0|\chi_j)}.
\end{equation}
Here the brackets indicate the average with respect to $p_0(z_0|\chi_j)$ over all $z_0\in\chi_j$.
Now we show that \eqreff{eq:s13} holds for all $\tau_k$, and Jarzynski's work $E_{\subj}(\chi_j)$ gives the difference between conformational free energy $G_{\obj}(\chi_j)\left(:=-\beta^{-1}\ln\int_{z\in\chi_j} e^{-\beta E_{\obj}(z)} \,dz\right)$ and $G(\chi_j)\left(:=-\beta^{-1}\ln\int_{z\in\chi_j} e^{-\beta E(z)} \,dz\right)$.

To this end, we may write $p(\chi_j,\tau_k)$ in \eqreff{eq:s3} by taking views II and I, respectively, as follows:
\begin{equation}\label{eq:s15}
p(\chi_j,\tau_k) = \frac{e^{-\beta G_{\obj}(\chi_j)}}{e^{-\beta\Psi_{\rm II}(\chi_j,\tau_k)}}=\frac{e^{-\beta G(\chi_j)}}{e^{-\beta\Psi_{\rm I}(\chi_j,\tau_k)}},
\end{equation}
which implies
\begin{equation}\label{eq:s16}
\Psi_{\rm I}(\chi_j,\tau_k)-\Psi_{\rm II}(\chi_j,\tau_k)=G(\chi_j)-G_{\obj}(\chi_j)
\end{equation}
for all $\tau_k$.  
At time $\tau_0$, \eqreff{eq:s16} implies  
\begin{equation}\label{eq:s17}
G(\chi_j)-G_{\obj}(\chi_j) = E_{\subj}(\chi_j)
\end{equation}
by \eqreff{eq:s13}.
Combining \eqreff{eq:s16} and \eqreff{eq:s17} shows that for all $\tau_k$,
\begin{equation}\label{eq:s18}
\Psi_{\rm I}(\chi_j,\tau_k)-\Psi_{\rm II}(\chi_j,\tau_k)= E_{\subj}(\chi_j), \end{equation}
proving that Jarzynski's work $E_{\subj}(\chi_j)$ links views I and II.

\subsection{Corollaries}
We have two corollaries.
Firstly, \eqreff{eq:s15} immediately implies the following fluctuation theorem for $\Psi$ that holds for all $\tau_k$: 
\begin{eqnarray}\label{eq:f}
\begin{aligned}
\left< e^{-\beta \Psi_{\rm II}(\chi_j,\tau_k)}\right>_{\chi_j} &= e^{-\beta A_0} \\
\left< e^{-\beta \Psi_{\rm I}(\chi_j,\tau_k)}\right>_{\chi_j} &= e^{-\beta A_1},
\end{aligned}
\end{eqnarray}
where $A_0\left(:=-\beta^{-1}\ln\sum_{\chi_j}\int_{z\in\chi_j}e^{-\beta E_{\obj}(z)} \,dz\right)$ and $A_1\left(:=-\beta^{-1}\ln\sum_{\chi_j}\int_{z\in\chi_j}e^{-\beta E(z)} \,dz\right)$ are the initial and the final equilibrium free energy, respectively. Here the brackets indicate the average over all $\chi_j$. By Jensen's inequality, \eqreff{eq:f} implies
\begin{eqnarray}
\begin{aligned}
\left< \Psi_{\rm II}(\chi_j,\tau_k)\right>_{\chi_j} &\ge  A_0 \\
\left< \Psi_{\rm I}(\chi_j,\tau_k)\right>_{\chi_j} &\ge A_1.
\end{aligned}
\end{eqnarray}

Secondly, the exponent $G_{\obj}(\chi_j)$ and $G(\chi_j)$ in \eqreff{eq:s15} does not imply that a system is in local equilibrium within $\chi_j$. We have the following non-equilibrium equalities for any $t$ 
\begin{eqnarray}\label{eq:g}
\begin{aligned}
\left< e^{-\beta \Psi_{\rm II}(z,t)}\delta_{\chi_j}(z) \right>_z &= e^{-\beta G_{\obj}(\chi_j)}  \\
\left< e^{-\beta \Psi_{\rm I}(z,t)}\delta_{\chi_j}(z) \right>_z & = e^{-\beta G(\chi_j)},
\end{aligned}
\end{eqnarray}
where the bracket indicates average over all $z$, and
$\delta_{\chi_j}(z)=1$ if $z\in\chi_j$ and $0$ 
otherwise. We remark that \eqreff{eq:g} links two known relations in the literature; one for work and the conformational free energy $G(\chi_j)$ \cite{hummer}, and the other for work and $\psi(z,t)$ \cite{local}.

\subsection{Extremal principle in view II}

The upper-bound of the average of $\Psi_{\rm II}(\chi_j,\tau_k)$ over all $\chi_j$ comes from equation (20) by taking view II as follows:
\begin{equation}\label{eq:s19}
\left<\Psi_{\rm II}(\chi_j,\tau_k)\right>_j \le A_0 + \left<W_{\subj}\right>,
\end{equation}
where $W_{\subj}:=E_{\subj}(z_0) - E_{\subj}(z_\tau)$ for a trajectory $\{z_t\}_{0\le t\le \tau}$, and the average $\left<W_{\subj}\right>$ is over all paths that end at time $\tau\in\tau_k$.
Now we convert this inequality into an equality by analyzing how the average internal-work, $\left<W_{\subj}\right>$, is consumed during the process.

By \eqreff{eq:s1}, we have
\begin{equation}\label{eq:s20}
\psi_{\rm I}(z,t) - \psi_{\rm II}(z,t) = E_{\subj}(z),
\end{equation}
which implies
\begin{equation}\label{eq:s21}
\left<W_{\subj}\right>= \Delta\left<\psi_{\rm II}\right>_z-\Delta\left< \psi_{\rm I}\right>_z,
\end{equation}
where $\Delta f:=f(\tau_k) - f(0)$ for any function $f$ of time. Here $f(\tau_k):=\left(\int_{\tau\in\tau_k}f(\tau) \,d\tau\right)/\delta \tau_k $ is the time average over all $\tau\in\tau_k$ so that we have
$\Delta\left<\psi_{\rm II}\right>_z := \left<\psi_{\rm II}(z,\tau)\right>_{p(z,\tau_k)} - \left<\psi_{\rm II}(z,0)\right>_{p(z,0)}$, and $\Delta\left<\psi_{\rm I}\right>_z := \left<\psi_{\rm I}(z,\tau)\right>_{p(z,\tau_k)} - \left<\psi_{\rm I}(z,0)\right>_{p(z,0)}$, where brackets $\left<\cdot\right>_{p(z,\tau_k)}$ indicate the average over all $z$ and $\tau\in\tau_k$ with respect to $p(z,\tau)$ (note that $\tau_0:=0$).
To engage $\Psi_{\rm II}(\chi_j,\tau_k)$ with \eqreff{eq:s21}, we consider $p(z,t | \chi_j,\tau_k)$ and the locally-equilibrated distribution $p_{\rm loc}^{\rm eq}(z,t | \chi_j,\tau_k)$ within $z\in\chi_j$, which reads $e^{-\beta E_{\obj}(z)}/e^{-\beta G_{\obj}(\chi_j)}$ in view II. Direct calculation of the Kullback-Leibler divergence $D\left(p(z,t |\chi_j,\tau_k)\parallel p_{\rm loc}^{\rm eq}(z,t | \chi_j,\tau_k)\right)$ using \eqreff{eq:s1} and \eqreff{eq:s3} gives
\begin{equation}\label{eq:s22}
\left<\psi_{\rm II}(z,t)\right>_{p(z,t|\chi_j,\tau_k)}= \Psi_{\rm II}(\chi_j,\tau_k)+D\left(p(z,t |\chi_j,\tau_k)\parallel p_{\rm loc}^{\rm eq}(z,t |\chi_j,\tau_k)\right).
\end{equation}
Taking the average of \eqreff{eq:s22} over all $\chi_j$ with respect to $p(\chi_j,t)$ gives
\begin{equation}\label{eq:s23} 
\left<\psi_{\rm II}(z,t)\right>_z = \left<\Psi_{\rm II}(\chi_j,\tau_k)\right>_j + D(p(z,t) \parallel p_{\rm loc}^{\rm eq}(z,t)).
\end{equation}
Finally, combing \eqreff{eq:s21} and \eqreff{eq:s23} proves
\begin{equation}\label{eq:s24} 
\left<W_{\subj}\right>=\Delta\left<\Psi_{\rm II}\right>_j-\Delta\left<\psi_{\rm I}\right>_z + \Delta I,
\end{equation}
where $\left<\Psi_{\rm II}\right>_j$ indicates the average of $\Psi_{\rm II}(\chi_j,\cdot)$ over all $\chi_j$, and $I(t):=D(p(z,t)\parallel p_{\rm loc}^{\rm eq}(z,t))$.  
\Eqreff{eq:s24} tells that some of the internal-work increases $\left<\Psi_{\rm II}\right>_j$,
some dissipate, resulting in the irreversible entropy production $-\Delta\left<\psi_{\rm I}\right>_z$,
and the rest develops local details away from local equilibrium
$D(p(z,\cdot)\parallel p_{\rm loc}^{\rm eq}(z,\cdot))$.

Now we return to the inequality for the upper-bound of $\left<\Psi_{\rm II}\right>_j$, \eqreff{eq:s19}, taking
$\tau_k\rightarrow\infty$, and then, we have
 \begin{equation}\label{eq:s25}
\left<\Psi_{\rm II}(\chi_j,\infty)\right>_j \le A_0 + \left<W_{\subj}\right>.
\end{equation}
During the process, the total internal-work becomes
\begin{equation}\label{eq:s26}
\left<W_{\subj}\right> = \left<E_{\subj}(z)\right>_{p_0(z)} - \left<E_{\subj}(z)\right>_{p_1(z)},
\end{equation}
and the irreversible entropy production $-\Delta\left<\psi_{\rm I}\right>_z$ reads:
\begin{equation}\label{eq:s27}
\left<\psi_{\rm I}(z,0)\right>_{p_0(z)} - \left<\psi_{\rm I}(z,\infty)\right>_{p_1(z)} = A_0+\left<E_{\subj}(z)\right>_{p_0(z)}    - A_1 
\end{equation}
by \eqreff{eq:s20}.
The amount of local-details developed away from local-equilibrium $\Delta I$ is  
\begin{equation}\label{eq:s28}
\left<\psi_{\rm II}(z,\infty)\right>_{p_1(z)} - \left<\Psi_{\rm II}(\chi_j,\infty)\right>_{p_1(\chi_j)} = \left<E_{\subj}(\chi_j)\right>_{p_1(\chi_j)} - \left<E_{\subj}(z)\right>_{p_1(z)}
\end{equation}
by \eqreff{eq:s18} and \eqreff{eq:s20} since we have $I(0)=0$.
We should subtract \eqreff{eq:s27} and \eqreff{eq:s28} from \eqreff{eq:s26} so that the maximally attainable value of $\left<\Psi_{\rm II}(\chi_j,\infty)\right>_j$ from \eqreff{eq:s25} reads
\begin{equation}
\left<\Psi_{\rm II}(\chi_j,\infty)\right>_j \le A_1 - \left<E_{\subj}(\chi_j)\right>_{p_1(\chi_j)}.
\end{equation}
Now \eqreff{eq:s18} tells that
\begin{equation}
\left<\Psi_{\rm II}(\chi_j,\infty)\right>_j = A_1 - \left<E_{\subj}(\chi_j)\right>_{p_1(\chi_j)},
\end{equation}
which proves that
\begin{equation}\nonumber
\left<\Psi_{\rm II}\right>_j \mbox{tends to be maximized}.
\end{equation}

\section{Simulation details}

\subsection{Activation of bi-stable potential}\label{sec5}
The initial distribution $p_0(\chi)$ for a hypothetical one-dimensional reaction coordinate $\chi (0\le\chi\le L)$ with $L=10$ is set to the uniform distribution, and at time $\tau_0$, we activated $G(\chi)$ which is a bistable potential represented in Fig. 3A(6).
We solved numerically $300,000$ times the over-damped Langevin equation:
$$\zeta \dot{\chi}=-\nabla E_{\subj}(\chi,t) + \xi,$$
where thermal fluctuation $\xi$ satisfies the fluctuation-dissipation relation $\langle\xi(t)\xi(t')\rangle=2k_{\rm B}T \zeta\delta(t-t')$, and $\zeta=1$.
We partitioned the one-dimensional domain into $50$ bins to form $\{\chi_j| j=1,\cdots, 50\}$ and counted the number of particles for each bin at each time to obtain $\Psi_{\rm I}(\chi_j,\tau_k)$, $\Psi_{\rm II}(\chi_j,\tau_k)$, $\left<\Psi_{\rm I}\right>_j$, and $\left<\Psi_{\rm II}\right>_j$ in Fig. 3.
After integrating over $4000$ steps using $0.01$ discretization interval, we deactivated $G(\chi)$ to simulate the inverse process and to calculate $\left<\Psi'_{\rm I}\right>_j$, and $\left<\Psi'_{\rm II}\right>_j$.

\subsection{Ribose binding protein}\label{sec:RBP}

We have analyzed the simulation results from Ravindranathan et al.~\cite{JMB2005} to calculate $\Psi_{\rm II}(\theta,\infty)$ and $G_{\obj}(\theta)$ in Fig. 4A by setting as follows:
\begin{equation*}
\begin{aligned}
E_{\obj}(z) &:= E_{\rm RBP}(x) + E_{\rm rib}(y)\\
E_{\subj}(z)&: = E_{\rm int}(x,y),
\end{aligned}
\end{equation*}
where $E_{\rm RBP}(x)$ and $E_{\rm rib}(y)$ are the internal energy of ribose-binding protein (RBP) and ribose, respectively, $E_{\rm int}(z)$ gives interaction between them, and $z=(x,y)$ is a phase-space point of the system.
A reaction coordinate is angle $\theta$ between the centers of mass of the two domains of RBP and the center of mass of the three-stranded hinge.
We consider the case that before time $\tau_0$, RBP and ribose are in equilibrium without interaction, and at time $\tau_0$, they start interaction by activating $E_{\subj}(z)$.

Ravindranathan et al.~\cite{JMB2005} carried out a total of 32 umbrella sampling molecular dynamics simulations \cite{umbrella1997}, 16 of RBP with ribose and 16 of RBP without ribose, using the OPLS-AA all-atom force field \cite{OPLS1996} with the AGBNP implicit solvent model \cite{solvent2004}. They took crystal structures of RBP from RCSB Protein Data Bank \cite{pdb2003}, added Hydrogen atoms, heated the system to 300 K over 3 ps, equilibrated for 225 ps at 300 K, and after equilibration, gathered data for 800 ps. The time-step is 1 fs, and all atoms are treated explicitly. They applied the weighted histogram analysis \cite{weighted1992,roux1995} to obtain the unbiased (independent of biasing potential) population distribution $p_0(\theta)$ of ribose-free RBP and $p_1(\theta)$ of ribose-bound RBP, which we have digitized to calculate $\beta(\Psi_{\rm II}(\theta, \infty)-A_0)$ by $\ln (p_1(\theta)/p_0(\theta))$, and $\beta(G_{\obj}(\theta)-A_0)$ by $-\ln p_0(\theta)$ for Fig 4A.

The use of data from ribose-free RBP simulations to obtain $p_0(\theta)$ for the complex of RBP and ribose with no interaction is justified since the conformational free energy of the complex can be decomposed into that of each molecules due to the absence of interactions between them as follows:  
\begin{eqnarray}
\begin{aligned}
p_0(\theta) &:= \frac{\int_{z\in\theta} e^{-\beta E_{\obj}(z)} dz}{\int_{z} e^{-\beta E_{\obj}(z)} dz}
=\frac{\int_{x\in\theta} e^{-\beta E_{\rm RBP}(x)} dx \int_{y} e^{-\beta E_{\rm rib}(y)} dy}
    {\int_{x} e^{-\beta E_{\rm RBP}(x)} dx \int_{y} e^{-\beta E_{\rm rib}(y)} dy}\\
  &=\frac{\int_{x\in\theta} e^{-\beta E_{\rm RBP}(x)} dx}{\int_{x} e^{-\beta E_{\rm RBP}(x)} dx}.
\end{aligned}
\end{eqnarray}


\begin{figure}[h]
\label{fig:figS1}
\includegraphics[width=16cm]{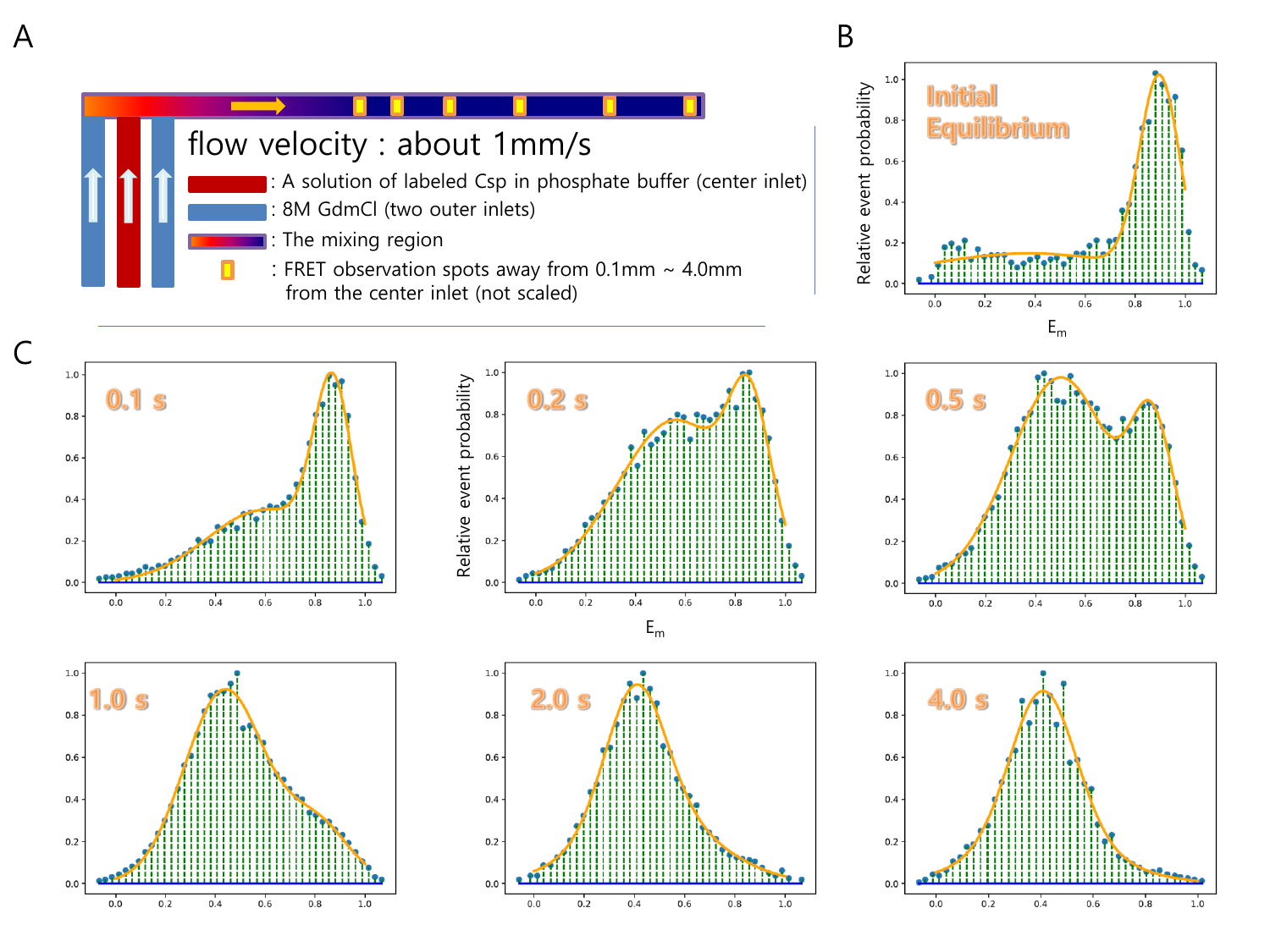}
\caption{
{\bf Microfluidic mixer and measured FRET efficiencies.} 
{\bf A,} Schematic representation of microfluidic mixer is drawn. Due to the flow velocity is about $1\mu m/ms$, the observation spots $0.1, 0.2, 0.5, 1.0, 2.0,$ and $4.0$~mm away from the center inlet correspond to the time delays of $\tau_k\approx 0.1, 0.2, 0.5, 1.0, 2.0,$ and $4.0$~s after an abrupt change of solution conditions at time $\tau_0=0$ to those that favor unfolding.
{\bf B,} The histogram of measured FRET efficiency $E_m$ in the initial (before mixing) solution conditions that favor folding. 
{\bf C,} Histograms of measured FRET efficiency $E_m$ at time $\tau_k$. In {\bf B} and {\bf C}, green dotted lines with blue solid-circle ends indicate the values of the relative event probability of $E_m$ digitized from \cite{lipman2003single} and orange curves are double Gaussian fits to the data.
}
\end{figure}

\subsection{Single-molecule FRET measurement}

Lipman et al. \cite{lipman2003single} have measured fluorescence resonance energy transfer (FRET) efficiencies of single molecules under conditions far from equilibrium by coupling a microfabricated laminar-flow mixer, which enables time-resolved measurement of FRET after an abrupt change in solution conditions. They have carried out the kinetic experiments for the folding of the cold shock protein (Csp) and the unfolding of Csp. We analyze their data for Csp unfolding.  

A solution of labeled Csp in phosphate buffer is prepared in the center inlet channel, and 8M GdmCl molecules are in the two outer inlet channels (see Fig. 5 in the Appendix). Solutions in the inlets are driven to the mixing region with compressed air and are mixed by the time they have arrived a point 50~$\mu$m beyond the center inlet, which corresponds to 50~ms due to the flow velocity of about 1~$\mu$m/ms. Actual FRET measurements are made at distances 
$\ge 100$~$\mu$m from the center inlet, which corresponds to $\ge 0.1$~s.

We digitized histograms of measured FRET efficiency $E_m$ during unfolding and fitted double-Gaussians to the data, which is justified since Csp exhibits two-state folding/unfolding kinetics \cite{perl1998,wassenberg1999}, as shown in Fig. 5 in Appendix (we refer the reader to \cite{lipman2003single} for more details on obtaining $E_m$). We observe that as the time flows, the event probability at lower transfer efficiency (greater dye-dye separation) increases, indicating that non-equilibrium unfolding progresses. We calculated $\beta(\Psi_{\rm II}(E_m, \tau_k)-A_0)$ by $\ln (p(E_m,\tau_k)/p_0(E_m))$, and $\beta(G_{\obj}(E_m)-A_0)$ by $-\ln p_0(E_m)$ for Figs 4B and 4C.

The use of data from the initial solution condition without GdmCl to obtain $p_0(E_m)$ for the complex of Csp with inert GdmCl is justified again since the conformational free energy of the complex can be decomposed into that of the solution of Csp and that of GdmCl molecules due to the absence of interactions between them as follows:  
\begin{eqnarray}
\begin{aligned}
p_0(E_m) &:= \frac{\int_{z\in E_m} e^{-\beta E_{\obj}(z)} dz}{\int_{z} e^{-\beta E_{\obj}(z)} dz}
=\frac{\int_{x\in E_m} e^{-\beta E_{\rm CspInBuffer}(x)} dx \int_{y} e^{-\beta E_{\rm GdmCl}(y)} dy}
    {\int_{x} e^{-\beta E_{\rm CspInBuffer}(x)} dx \int_{y} e^{-\beta E_{\rm GdmCl}(y)} dy}\\
  &=\frac{\int_{x\in E_m} e^{-\beta E_{\rm CspInBuffer}(x)} dx}{\int_{x} e^{-\beta E_{\rm CspInBuffer}(x)} dx},
\end{aligned}
\end{eqnarray}
where the last term corresponds to the relative event probability shown in Fig. 5 in Appendix.

\bibliographystyle{naturemag_noURL}
\bibliography{bib}

\begin{thebibliography}{10}
\expandafter\ifx\csname url\endcsname\relax
  \def\url#1{\texttt{#1}}\fi
\expandafter\ifx\csname urlprefix\endcsname\relax\def\urlprefix{URL }\fi
\providecommand{\bibinfo}[2]{#2}
\providecommand{\eprint}[2][]{\url{#2}}

\bibitem{anfinsen1975}
\bibinfo{author}{Anfinsen, C.~B.} \& \bibinfo{author}{Scheraga, H.~A.}
\newblock \bibinfo{title}{{Experimental and theoretical aspects of protein
  folding}}.
\newblock \emph{\bibinfo{journal}{Advances in protein chemistry}}
  \textbf{\bibinfo{volume}{29}}, \bibinfo{pages}{205--300}
  (\bibinfo{year}{1975}).

\bibitem{shape}
\bibinfo{author}{Robertson, E.~G.} \& \bibinfo{author}{Simons, J.~P.}
\newblock \bibinfo{title}{{Getting into shape: conformational and
  supramolecular landscapes in small biomolecules and their hydrated
  clusters}}.
\newblock \emph{\bibinfo{journal}{Physical Chemistry Chemical Physics}}
  \textbf{\bibinfo{volume}{3}}, \bibinfo{pages}{1--18} (\bibinfo{year}{2001}).

\bibitem{molecular}
\bibinfo{author}{Fan, E.}, \bibinfo{author}{{Van Arman}, S.~A.},
  \bibinfo{author}{Kincaid, S.} \& \bibinfo{author}{Hamilton, A.~D.}
\newblock \bibinfo{title}{{Molecular recognition: hydrogen-bonding receptors
  that function in highly competitive solvents}}.
\newblock \emph{\bibinfo{journal}{Journal of the American Chemical Society}}
  \textbf{\bibinfo{volume}{115}}, \bibinfo{pages}{369--370}
  (\bibinfo{year}{1993}).

\bibitem{ham2012pnas}
\bibinfo{author}{Chong, S.-H.} \& \bibinfo{author}{Ham, S.}
\newblock \bibinfo{title}{{Impact of chemical heterogeneity on protein
  self-assembly in water}}.
\newblock \emph{\bibinfo{journal}{Proceedings of the National Academy of
  Sciences}} \textbf{\bibinfo{volume}{109}}, \bibinfo{pages}{7636--7641}
  (\bibinfo{year}{2012}).

\bibitem{funnel1991}
\bibinfo{author}{Frauenfelder, H.}, \bibinfo{author}{Sligar, S.~G.} \&
  \bibinfo{author}{Wolynes, P.~G.}
\newblock \bibinfo{title}{{The energy landscapes and motions of proteins}}.
\newblock \emph{\bibinfo{journal}{Urbana}} \textbf{\bibinfo{volume}{51}},
  \bibinfo{pages}{61801} (\bibinfo{year}{1991}).

\bibitem{funnel1992}
\bibinfo{author}{Leopold, P.~E.}, \bibinfo{author}{Montal, M.} \&
  \bibinfo{author}{Onuchic, J.~N.}
\newblock \bibinfo{title}{{Protein folding funnels: a kinetic approach to the
  sequence-structure relationship.}}
\newblock \emph{\bibinfo{journal}{Proceedings of the National Academy of
  Sciences}} \textbf{\bibinfo{volume}{89}}, \bibinfo{pages}{8721--8725}
  (\bibinfo{year}{1992}).

\bibitem{funnel1997}
\bibinfo{author}{Lazaridis, T.} \& \bibinfo{author}{Karplus, M.}
\newblock \bibinfo{title}{{" New view" of protein folding reconciled with the
  old through multiple unfolding simulations}}.
\newblock \emph{\bibinfo{journal}{Science}} \textbf{\bibinfo{volume}{278}},
  \bibinfo{pages}{1928--1931} (\bibinfo{year}{1997}).

\bibitem{revSears2008}
\bibinfo{author}{{E.M. Sevick}}, \bibinfo{author}{{R. Prabhakar}},
  \bibinfo{author}{{Stephen R. Williams}} \& \bibinfo{author}{Searles, D.~J.}
\newblock \bibinfo{title}{{Fluctuation Theorems}}.
\newblock \emph{\bibinfo{journal}{Annual Review of Physical Chemistry}}
  \textbf{\bibinfo{volume}{59}}, \bibinfo{pages}{603--633}
  (\bibinfo{year}{2008}).

\bibitem{jarReview}
\bibinfo{author}{Jarzynski, C.}
\newblock \bibinfo{title}{{Equalities and inequalities: Irreversibility and the
  second law of thermodynamics at the nanoscale}}.
\newblock \emph{\bibinfo{journal}{Annu. Rev. Codens. Matter Phys.}}
  \textbf{\bibinfo{volume}{2}}, \bibinfo{pages}{329--351}
  (\bibinfo{year}{2011}).

\bibitem{revSeifert}
\bibinfo{author}{Seifert, U.}
\newblock \bibinfo{title}{{Stochastic thermodynamics, fluctuation theorems and
  molecular machines}}.
\newblock \emph{\bibinfo{journal}{Rep. Prog. Phys.}}
  \textbf{\bibinfo{volume}{75}}, \bibinfo{pages}{126001}
  (\bibinfo{year}{2012}).

\bibitem{liph2001}
\bibinfo{author}{Liphardt, J.}, \bibinfo{author}{Onoa, B.},
  \bibinfo{author}{Smith, S.~B.}, \bibinfo{author}{Tinoco, I.} \&
  \bibinfo{author}{Bustamante, C.}
\newblock \bibinfo{title}{{Reversible unfolding of single RNA molecules by
  mechanical force}}.
\newblock \emph{\bibinfo{journal}{Science}} \textbf{\bibinfo{volume}{292}},
  \bibinfo{pages}{733--737} (\bibinfo{year}{2001}).

\bibitem{expColin}
\bibinfo{author}{Collin, D.} \emph{et~al.}
\newblock \bibinfo{title}{{Verification of the Crooks fluctuation theorem and
  recovery of RNA folding free energies}}.
\newblock \emph{\bibinfo{journal}{Nature}} \textbf{\bibinfo{volume}{437}},
  \bibinfo{pages}{231--234} (\bibinfo{year}{2005}).

\bibitem{ritort2012}
\bibinfo{author}{Alemany, A.}, \bibinfo{author}{Mossa, A.},
  \bibinfo{author}{Junier, I.} \& \bibinfo{author}{Ritort, F.}
\newblock \bibinfo{title}{{Experimental free-energy measurements of kinetic
  molecular states using fluctuation theorems}}.
\newblock \emph{\bibinfo{journal}{Nature Phys.}}  (\bibinfo{year}{2012}).

\bibitem{personality2007}
\bibinfo{author}{Henzler-Wildman, K.} \& \bibinfo{author}{Kern, D.}
\newblock \bibinfo{title}{{Dynamic personalities of proteins}}.
\newblock \emph{\bibinfo{journal}{Nature}} \textbf{\bibinfo{volume}{450}},
  \bibinfo{pages}{964} (\bibinfo{year}{2007}).

\bibitem{hyeon2012}
\bibinfo{author}{Hyeon, C.}, \bibinfo{author}{Lee, J.}, \bibinfo{author}{Yoon,
  J.}, \bibinfo{author}{Hohng, S.} \& \bibinfo{author}{Thirumalai, D.}
\newblock \bibinfo{title}{{Hidden complexity in the isomerization dynamics of
  Holliday junctions}}.
\newblock \emph{\bibinfo{journal}{Nature chemistry}}
  \textbf{\bibinfo{volume}{4}}, \bibinfo{pages}{907--914}
  (\bibinfo{year}{2012}).

\bibitem{roldan2014}
\bibinfo{author}{Rold{\'{a}}n, {\'{E}}.}, \bibinfo{author}{Martinez, I.~A.},
  \bibinfo{author}{Parrondo, J. M.~R.} \& \bibinfo{author}{Petrov, D.}
\newblock \bibinfo{title}{{Universal features in the energetics of symmetry
  breaking}}.
\newblock \emph{\bibinfo{journal}{Nature Physics}}  (\bibinfo{year}{2014}).

\bibitem{multiple2010}
\bibinfo{author}{Solomatin, S.~V.}, \bibinfo{author}{Greenfeld, M.},
  \bibinfo{author}{Chu, S.} \& \bibinfo{author}{Herschlag, D.}
\newblock \bibinfo{title}{{Multiple native states reveal persistent ruggedness
  of an RNA folding landscape}}.
\newblock \emph{\bibinfo{journal}{Nature}} \textbf{\bibinfo{volume}{463}},
  \bibinfo{pages}{681} (\bibinfo{year}{2010}).

\bibitem{hill1962}
\bibinfo{author}{Hill, T.~L.}
\newblock \bibinfo{title}{{Thermodynamics of small systems}}.
\newblock \emph{\bibinfo{journal}{The Journal of Chemical Physics}}
  \textbf{\bibinfo{volume}{36}}, \bibinfo{pages}{3182--3197}
  (\bibinfo{year}{1962}).

\bibitem{hill2012}
\bibinfo{author}{Hill, T.}
\newblock \emph{\bibinfo{title}{{Free energy transduction in biology: the
  steady-state kinetic and thermodynamic formalism}}}
  (\bibinfo{publisher}{Elsevier}, \bibinfo{year}{2012}).

\bibitem{parmeggiani1999}
\bibinfo{author}{Parmeggiani, A.}, \bibinfo{author}{J{\"{u}}licher, F.},
  \bibinfo{author}{Ajdari, A.} \& \bibinfo{author}{Prost, J.}
\newblock \bibinfo{title}{{Energy transduction of isothermal ratchets: Generic
  aspects and specific examples close to and far from equilibrium}}.
\newblock \emph{\bibinfo{journal}{Physical Review E}}
  \textbf{\bibinfo{volume}{60}}, \bibinfo{pages}{2127} (\bibinfo{year}{1999}).

\bibitem{fisher1999}
\bibinfo{author}{Fisher, M.~E.} \& \bibinfo{author}{Kolomeisky, A.~B.}
\newblock \bibinfo{title}{{The force exerted by a molecular motor}}.
\newblock \emph{\bibinfo{journal}{Proceedings of the National Academy of
  Sciences}} \textbf{\bibinfo{volume}{96}}, \bibinfo{pages}{6597--6602}
  (\bibinfo{year}{1999}).

\bibitem{fisher2001}
\bibinfo{author}{Fisher, M.~E.} \& \bibinfo{author}{Kolomeisky, A.~B.}
\newblock \bibinfo{title}{{Simple mechanochemistry describes the dynamics of
  kinesin molecules}}.
\newblock \emph{\bibinfo{journal}{Proceedings of the National Academy of
  Sciences}} \textbf{\bibinfo{volume}{98}}, \bibinfo{pages}{7748--7753}
  (\bibinfo{year}{2001}).

\bibitem{bustamante2001}
\bibinfo{author}{Bustamante, C.}, \bibinfo{author}{Keller, D.} \&
  \bibinfo{author}{Oster, G.}
\newblock \bibinfo{title}{{The physics of molecular motors}}.
\newblock \emph{\bibinfo{journal}{Accounts of Chemical Research}}
  \textbf{\bibinfo{volume}{34}}, \bibinfo{pages}{412--420}
  (\bibinfo{year}{2001}).

\bibitem{schmiedl2008}
\bibinfo{author}{Schmiedl, T.} \& \bibinfo{author}{Seifert, U.}
\newblock \bibinfo{title}{{Efficiency of molecular motors at maximum power}}.
\newblock \emph{\bibinfo{journal}{EPL (Europhysics Letters)}}
  \textbf{\bibinfo{volume}{83}}, \bibinfo{pages}{30005} (\bibinfo{year}{2008}).

\bibitem{liepelt2007}
\bibinfo{author}{Liepelt, S.} \& \bibinfo{author}{Lipowsky, R.}
\newblock \bibinfo{title}{{Kinesin's network of chemomechanical motor cycles}}.
\newblock \emph{\bibinfo{journal}{Physical review letters}}
  \textbf{\bibinfo{volume}{98}}, \bibinfo{pages}{258102}
  (\bibinfo{year}{2007}).

\bibitem{hwang2016}
\bibinfo{author}{Hwang, W.} \& \bibinfo{author}{Hyeon, C.}
\newblock \bibinfo{title}{{Quantifying the heat dissipation from a molecular
  motor's transport properties in nonequilibrium steady states}}.
\newblock \emph{\bibinfo{journal}{The Journal of Physical Chemistry Letters}}
  \textbf{\bibinfo{volume}{8}}, \bibinfo{pages}{250--256}
  (\bibinfo{year}{2016}).

\bibitem{hwang2018}
\bibinfo{author}{Hwang, W.} \& \bibinfo{author}{Hyeon, C.}
\newblock \bibinfo{title}{{Energetic Costs, Precision, and Transport Efficiency
  of Molecular Motors}}.
\newblock \emph{\bibinfo{journal}{The journal of physical chemistry letters}}
  \textbf{\bibinfo{volume}{9}}, \bibinfo{pages}{513--520}
  (\bibinfo{year}{2018}).

\bibitem{sasa}
\bibinfo{author}{Hatano, T.} \& \bibinfo{author}{Sasa, S.-i.}
\newblock \bibinfo{title}{{Steady-state thermodynamics of Langevin systems}}.
\newblock \emph{\bibinfo{journal}{Phys. Rev. Lett.}}
  \textbf{\bibinfo{volume}{86}}, \bibinfo{pages}{3463--3466}
  (\bibinfo{year}{2001}).

\bibitem{seifert05}
\bibinfo{author}{Seifert, U.}
\newblock \bibinfo{title}{{Entropy production along a stochastic trajectory and
  an integral fluctuation theorem}}.
\newblock \emph{\bibinfo{journal}{Phys. Rev. Lett.}}
  \textbf{\bibinfo{volume}{95}}, \bibinfo{pages}{40602} (\bibinfo{year}{2005}).

\bibitem{bochkov1977}
\bibinfo{author}{Bochkov, G.~N.} \& \bibinfo{author}{Kuzovlev, Y.~E.}
\newblock \bibinfo{title}{{General theory of thermal fluctuations in nonlinear
  systems}}.
\newblock \emph{\bibinfo{journal}{Zh. Eksp. Teor. Fiz}}
  \textbf{\bibinfo{volume}{72}}, \bibinfo{pages}{238--243}
  (\bibinfo{year}{1977}).

\bibitem{bochkov1979}
\bibinfo{author}{Bochkov, G.~N.} \& \bibinfo{author}{Kuzovlev, Y.~E.}
\newblock \bibinfo{title}{{Fluctuation-dissipation relations for nonequilibrium
  processes in open systems}}.
\newblock \emph{\bibinfo{journal}{Soviet Journal of Experimental and
  Theoretical Physics}} \textbf{\bibinfo{volume}{49}}, \bibinfo{pages}{543}
  (\bibinfo{year}{1979}).

\bibitem{jar}
\bibinfo{author}{Jarzynski, C.}
\newblock \bibinfo{title}{{Nonequilibrium equality for free energy
  differences}}.
\newblock \emph{\bibinfo{journal}{Phys. Rev. Lett.}}
  \textbf{\bibinfo{volume}{78}}, \bibinfo{pages}{2690--2693}
  (\bibinfo{year}{1997}).

\bibitem{crooks99}
\bibinfo{author}{Crooks, G.~E.}
\newblock \bibinfo{title}{{Entropy production fluctuation theorem and the
  nonequilibrium work relation for free energy differences}}.
\newblock \emph{\bibinfo{journal}{Phys. Rev. E}} \textbf{\bibinfo{volume}{60}},
  \bibinfo{pages}{2721--2726} (\bibinfo{year}{1999}).

\bibitem{local}
\bibinfo{author}{Jinwoo, L.} \& \bibinfo{author}{Tanaka, H.}
\newblock \bibinfo{title}{{Local non-equilibrium thermodynamics}}.
\newblock \emph{\bibinfo{journal}{Sci.Rep.}} \textbf{\bibinfo{volume}{5}},
  \bibinfo{pages}{7832} (\bibinfo{year}{2015}).

\bibitem{jar2007work}
\bibinfo{author}{Jarzynski, C.}
\newblock \bibinfo{title}{{Comparison of far-from-equilibrium work relations}}.
\newblock \emph{\bibinfo{journal}{C. R. Physique}}
  \textbf{\bibinfo{volume}{8}}, \bibinfo{pages}{495--506}
  (\bibinfo{year}{2007}).

\bibitem{campisi2011}
\bibinfo{author}{Campisi, M.}, \bibinfo{author}{H{\"{a}}nggi, P.} \&
  \bibinfo{author}{Talkner, P.}
\newblock \bibinfo{title}{{Colloquium: Quantum fluctuation relations:
  Foundations and applications}}.
\newblock \emph{\bibinfo{journal}{Reviews of Modern Physics}}
  \textbf{\bibinfo{volume}{83}}, \bibinfo{pages}{771} (\bibinfo{year}{2011}).

\bibitem{liph2002}
\bibinfo{author}{Liphardt, J.}, \bibinfo{author}{Dumont, S.},
  \bibinfo{author}{Smith, S.~B.}, \bibinfo{author}{{Tinoco Jr}, I.} \&
  \bibinfo{author}{Bustamante, C.}
\newblock \bibinfo{title}{{Equilibrium information from nonequilibrium
  measurements in an experimental test of Jarzynski's equality}}.
\newblock \emph{\bibinfo{journal}{Science}} \textbf{\bibinfo{volume}{296}},
  \bibinfo{pages}{1832--1835} (\bibinfo{year}{2002}).

\bibitem{expSasa}
\bibinfo{author}{Trepagnier, E.~H.} \emph{et~al.}
\newblock \bibinfo{title}{{Experimental test of Hatano and Sasa's
  nonequilibrium steady-state equality}}.
\newblock \emph{\bibinfo{journal}{Proc. Nat. Acad. Sci. USA}}
  \textbf{\bibinfo{volume}{101}}, \bibinfo{pages}{15038--15041}
  (\bibinfo{year}{2004}).

\bibitem{kur}
\bibinfo{author}{Kurchan, J.}
\newblock \bibinfo{title}{{No Title}}.
\newblock \emph{\bibinfo{journal}{J. Phys. A: Math. Gen.}}
  \textbf{\bibinfo{volume}{31}}, \bibinfo{pages}{3719} (\bibinfo{year}{1998}).

\bibitem{maes1999}
\bibinfo{author}{Maes, C.}
\newblock \bibinfo{title}{{The fluctuation theorem as a Gibbs property}}.
\newblock \emph{\bibinfo{journal}{Journal of statistical physics}}
  \textbf{\bibinfo{volume}{95}}, \bibinfo{pages}{367--392}
  (\bibinfo{year}{1999}).

\bibitem{jar2000}
\bibinfo{author}{Jarzynski, C.}
\newblock \bibinfo{title}{{Hamiltonian derivation of a detailed fluctuation
  theorem}}.
\newblock \emph{\bibinfo{journal}{J. Stat. Phys.}}
  \textbf{\bibinfo{volume}{98}}, \bibinfo{pages}{77--102}
  (\bibinfo{year}{2000}).

\bibitem{pdb2003}
\bibinfo{author}{Berman, H.~M.}, \bibinfo{author}{Bourne, P.~E.},
  \bibinfo{author}{Westbrook, J.} \& \bibinfo{author}{Zardecki, C.}
\newblock \bibinfo{title}{The protein data bank}.
\newblock In \emph{\bibinfo{booktitle}{Protein Structure}},
  \bibinfo{pages}{394--410} (\bibinfo{publisher}{CRC Press},
  \bibinfo{year}{2003}).

\bibitem{balian}
\bibinfo{author}{Balian, R.}
\newblock \emph{\bibinfo{title}{{From Microphysics to Macrophysics: Methods and
  Applications of Statistical Physics. Volume I (Theoretical and Mathematical
  Physics)}}} (\bibinfo{publisher}{Springer}, \bibinfo{year}{2006}).

\bibitem{jar2013prx}
\bibinfo{author}{Deffner, S.} \& \bibinfo{author}{Jarzynski, C.}
\newblock \bibinfo{title}{{Information processing and the second law of
  thermodynamics: An inclusive, Hamiltonian approach}}.
\newblock \emph{\bibinfo{journal}{Physical Review X}}
  \textbf{\bibinfo{volume}{3}}, \bibinfo{pages}{41003} (\bibinfo{year}{2013}).

\bibitem{seifert2011stochastic}
\bibinfo{author}{Seifert, U.}
\newblock \bibinfo{title}{{Stochastic thermodynamics of single enzymes and
  molecular motors}}.
\newblock \emph{\bibinfo{journal}{The European Physical Journal E}}
  \textbf{\bibinfo{volume}{34}}, \bibinfo{pages}{26} (\bibinfo{year}{2011}).

\bibitem{mesoscopic}
\bibinfo{author}{Reguera, D.}, \bibinfo{author}{Rub$\backslash$i, J.~M.} \&
  \bibinfo{author}{Vilar, J. M.~G.}
\newblock \bibinfo{title}{{The Mesoscopic Dynamics of Thermodynamic Systems}}.
\newblock \emph{\bibinfo{journal}{J. Phys. Chem. B}}
  \textbf{\bibinfo{volume}{109}}, \bibinfo{pages}{21502--21515}
  (\bibinfo{year}{2005}).

\bibitem{qian2001relative}
\bibinfo{author}{Qian, H.}
\newblock \bibinfo{title}{{Relative entropy: Free energy associated with
  equilibrium fluctuations and nonequilibrium deviations}}.
\newblock \emph{\bibinfo{journal}{Physical Review E}}
  \textbf{\bibinfo{volume}{63}}, \bibinfo{pages}{42103} (\bibinfo{year}{2001}).

\bibitem{JMB2005}
\bibinfo{author}{Ravindranathan, K.~P.}, \bibinfo{author}{Gallicchio, E.} \&
  \bibinfo{author}{Levy, R.~M.}
\newblock \bibinfo{title}{{Conformational equilibria and free energy profiles
  for the allosteric transition of the ribose-binding protein}}.
\newblock \emph{\bibinfo{journal}{Journal of molecular biology}}
  \textbf{\bibinfo{volume}{353}}, \bibinfo{pages}{196--210}
  (\bibinfo{year}{2005}).

\bibitem{lipman2003single}
\bibinfo{author}{Lipman, E.~A.}, \bibinfo{author}{Schuler, B.},
  \bibinfo{author}{Bakajin, O.} \& \bibinfo{author}{Eaton, W.~A.}
\newblock \bibinfo{title}{Single-molecule measurement of protein folding
  kinetics}.
\newblock \emph{\bibinfo{journal}{Science}} \textbf{\bibinfo{volume}{301}},
  \bibinfo{pages}{1233--1235} (\bibinfo{year}{2003}).

\bibitem{interactions2007}
\bibinfo{author}{O'Brien, E.~P.}, \bibinfo{author}{Dima, R.~I.},
  \bibinfo{author}{Brooks, B.} \& \bibinfo{author}{Thirumalai, D.}
\newblock \bibinfo{title}{{Interactions between hydrophobic and ionic solutes
  in aqueous guanidinium chloride and urea solutions: lessons for protein
  denaturation mechanism}}.
\newblock \emph{\bibinfo{journal}{Journal of the American Chemical Society}}
  \textbf{\bibinfo{volume}{129}}, \bibinfo{pages}{7346--7353}
  (\bibinfo{year}{2007}).

\bibitem{precision2018}
\bibinfo{author}{Hellenkamp, B.} \emph{et~al.}
\newblock \bibinfo{title}{Precision and accuracy of single-molecule fret
  measurements?a multi-laboratory benchmark study}.
\newblock \emph{\bibinfo{journal}{Nature methods}}
  \textbf{\bibinfo{volume}{15}}, \bibinfo{pages}{669} (\bibinfo{year}{2018}).

\bibitem{roy2008practical}
\bibinfo{author}{Roy, R.}, \bibinfo{author}{Hohng, S.} \& \bibinfo{author}{Ha,
  T.}
\newblock \bibinfo{title}{A practical guide to single-molecule fret}.
\newblock \emph{\bibinfo{journal}{Nature methods}}
  \textbf{\bibinfo{volume}{5}}, \bibinfo{pages}{507} (\bibinfo{year}{2008}).

\bibitem{joo2008advances}
\bibinfo{author}{Joo, C.}, \bibinfo{author}{Balci, H.},
  \bibinfo{author}{Ishitsuka, Y.}, \bibinfo{author}{Buranachai, C.} \&
  \bibinfo{author}{Ha, T.}
\newblock \bibinfo{title}{Advances in single-molecule fluorescence methods for
  molecular biology}.
\newblock \emph{\bibinfo{journal}{Annu. Rev. Biochem.}}
  \textbf{\bibinfo{volume}{77}}, \bibinfo{pages}{51--76}
  (\bibinfo{year}{2008}).

\bibitem{schuler2008protein}
\bibinfo{author}{Schuler, B.} \& \bibinfo{author}{Eaton, W.~A.}
\newblock \bibinfo{title}{Protein folding studied by single-molecule fret}.
\newblock \emph{\bibinfo{journal}{Current opinion in structural biology}}
  \textbf{\bibinfo{volume}{18}}, \bibinfo{pages}{16--26}
  (\bibinfo{year}{2008}).

\bibitem{jar_lag2009}
\bibinfo{author}{Vaikuntanathan, S.} \& \bibinfo{author}{Jarzynski, C.}
\newblock \bibinfo{title}{{Dissipation and lag in irreversible processes}}.
\newblock \emph{\bibinfo{journal}{Europhys. Lett.}}
  \textbf{\bibinfo{volume}{87}}, \bibinfo{pages}{60005} (\bibinfo{year}{2009}).

\bibitem{hummer}
\bibinfo{author}{Hummer, G.} \& \bibinfo{author}{Szabo, A.}
\newblock \bibinfo{title}{{Free energy reconstruction from nonequilibrium
  single-molecule pulling experiments}}.
\newblock \emph{\bibinfo{journal}{Proc. Nat. Acad. Sci. USA}}
  \textbf{\bibinfo{volume}{98}}, \bibinfo{pages}{3658--3661}
  (\bibinfo{year}{2001}).

\bibitem{umbrella1997}
\bibinfo{author}{Bartels, C.} \& \bibinfo{author}{Karplus, M.}
\newblock \bibinfo{title}{{Multidimensional adaptive umbrella sampling:
  applications to main chain and side chain peptide conformations}}.
\newblock \emph{\bibinfo{journal}{Journal of Computational Chemistry}}
  \textbf{\bibinfo{volume}{18}}, \bibinfo{pages}{1450--1462}
  (\bibinfo{year}{1997}).

\bibitem{OPLS1996}
\bibinfo{author}{Jorgensen, W.~L.}, \bibinfo{author}{Maxwell, D.~S.} \&
  \bibinfo{author}{Tirado-Rives, J.}
\newblock \bibinfo{title}{{Development and testing of the OPLS all-atom force
  field on conformational energetics and properties of organic liquids}}.
\newblock \emph{\bibinfo{journal}{J. Am. Chem. Soc}}
  \textbf{\bibinfo{volume}{118}}, \bibinfo{pages}{11225--11236}
  (\bibinfo{year}{1996}).

\bibitem{solvent2004}
\bibinfo{author}{Gallicchio, E.} \& \bibinfo{author}{Levy, R.~M.}
\newblock \bibinfo{title}{{AGBNP: An analytic implicit solvent model suitable
  for molecular dynamics simulations and high-resolution modeling}}.
\newblock \emph{\bibinfo{journal}{Journal of computational chemistry}}
  \textbf{\bibinfo{volume}{25}}, \bibinfo{pages}{479--499}
  (\bibinfo{year}{2004}).

\bibitem{weighted1992}
\bibinfo{author}{Kumar, S.}, \bibinfo{author}{Rosenberg, J.~M.},
  \bibinfo{author}{Bouzida, D.}, \bibinfo{author}{Swendsen, R.~H.} \&
  \bibinfo{author}{Kollman, P.~A.}
\newblock \bibinfo{title}{{The weighted histogram analysis method for
  free-energy calculations on biomolecules. I. The method}}.
\newblock \emph{\bibinfo{journal}{Journal of computational chemistry}}
  \textbf{\bibinfo{volume}{13}}, \bibinfo{pages}{1011--1021}
  (\bibinfo{year}{1992}).

\bibitem{roux1995}
\bibinfo{author}{Roux, B.}
\newblock \bibinfo{title}{{The calculation of the potential of mean force using
  computer simulations}}.
\newblock \emph{\bibinfo{journal}{Computer physics communications}}
  \textbf{\bibinfo{volume}{91}}, \bibinfo{pages}{275--282}
  (\bibinfo{year}{1995}).

\bibitem{perl1998}
\bibinfo{author}{Perl, D.} \emph{et~al.}
\newblock \bibinfo{title}{Conservation of rapid two-state folding in
  mesophilic, thermophilic and hyperthermophilic cold shock proteins}.
\newblock \emph{\bibinfo{journal}{Nature structural biology}}
  \textbf{\bibinfo{volume}{5}}, \bibinfo{pages}{229} (\bibinfo{year}{1998}).

\bibitem{wassenberg1999}
\bibinfo{author}{Wassenberg, D.}, \bibinfo{author}{Welker, C.} \&
  \bibinfo{author}{Jaenicke, R.}
\newblock \bibinfo{title}{Thermodynamics of the unfolding of the cold-shock
  protein from thermotoga maritima}.
\newblock \emph{\bibinfo{journal}{Journal of molecular biology}}
  \textbf{\bibinfo{volume}{289}}, \bibinfo{pages}{187--193}
  (\bibinfo{year}{1999}).

\end{thebibliography}

\end{document}